\pdfoutput=1
\documentclass[ALICE,manyauthors]{cernphprep}
\usepackage[comma,square,numbers,sort&compress]{natbib}

\usepackage[export]{adjustbox}
\usepackage{hyperref}
\usepackage{color}
\usepackage{lineno}
\usepackage{booktabs}
\usepackage{upgreek}

\newcommand{\pA}{{\mbox{pA}}\xspace}
\newcommand{\pai}{$\pi^{0}$\xspace}
\newcommand{\paitable}{$\pi^{\mbox{\tiny 0}}$\xspace}
\newcommand{\e}{$\eta$\xspace}
\newcommand{\etapai}{$\eta/\pi^{0}$\xspace}
\newcommand{\pT}{\ensuremath{p_{\mbox{\tiny T}}}\xspace}
\newcommand{\sNN}{$\sqrt{s_{\mbox{\tiny NN}}}$\xspace~=~2.76~TeV\xspace}
\newcommand{\sNNR}{$\sqrt{s_{\mbox{\tiny NN}}}$\xspace~=~200~GeV\xspace}

\newcommand{\mT}{$m_{\mbox{\tiny T}}$\xspace}

\newcommand{\Pb}{{\mbox{Pb--Pb}}\xspace}
\newcommand{\AACol}{{\mbox{AA}}\xspace}

\newcommand{\MeVc}{MeV/$c$\xspace}
\newcommand{\GeVc}{GeV/$c$\xspace}
\newcommand{\massGeVc}{GeV/$c^{2}$\xspace}
\newcommand{\dEdx}{$\mbox{d}E/\mbox{d}x$\xspace}

\newcommand{\RAA}{\ensuremath{R_{\mbox{\tiny AA}}}\xspace}
\newcommand{\TAA}{\ensuremath{T_{\rm{AA}}}}

\begin{document}%

\begin{titlepage}
\PHyear{2018}
\PHnumber{040}      
\PHdate{14 March}   

%

\title{Neutral pion and \e meson production at mid-rapidity in \Pb collisions at \sNN}
\ShortTitle{Neutral pion and \e meson production at ALICE}   


\Collaboration{ALICE Collaboration\thanks{See Appendix~\ref{app:collab} for the list of collaboration members}}
\ShortAuthor{ALICE Collaboration} 

\begin{abstract}
Neutral pion and \e meson production in the transverse momentum range 1~$<$~\pT$<$~20~\GeVc 
have been measured at mid-rapidity by the ALICE experiment at the Large Hadron Collider 
(LHC) in central and semi-central \Pb collisions at \sNN. These results were obtained 
using the photon conversion method as well as the Photon Spectrometer (PHOS) and Electromagnetic Calorimeter (EMCal) detectors. The results 
extend the upper \pT reach of the previous ALICE \pai measurements from 12~\GeVc to 
20~\GeVc and present the first measurement of \e meson production in heavy-ion collisions 
at the LHC. The \etapai ratio is similar for the two centralities and reaches at high \pT 
a plateau value of 0.457~$\pm$~0.013$^{stat}$~$\pm$~0.018$^{syst}$. A suppression of similar 
magnitude for \pai and \e meson production is observed in \Pb collisions with respect to 
their production in pp collisions scaled by the number of binary nucleon-nucleon collisions. 
We discuss the results in terms of Next to Leading Order pQCD predictions and hydrodynamic models. The 
measurements show a stronger suppression than observed at lower center-of-mass energies 
in the \pT range 6~$<$~\pT$<$~10~\GeVc. For \pT $<$~3~\GeVc, hadronization models describe 
the \pai results while for the  \e some tension is observed.
\end{abstract}
 
\end{titlepage}
\setcounter{page}{2}

\section{Introduction}\label{sectionintro}

Quantum Chromodynamics (QCD)~\cite{Patrignani:2016xqp}, the fundamental theory of strong 
interactions, predicts that, above a certain critical energy density, hadrons melt into a 
Quark-Gluon Plasma (QGP)~\cite{Ding:2015ona,Braun-Munzinger:2015hba}. Such a state of matter 
is believed to have existed a few microseconds after the Big Bang~\cite{Boyanovsky:2006bf}. 
One of the goals of lattice QCD calculations is the understanding of the properties of 
strongly interacting matter and the nature of the phase transition that depends on the 
values of the quark masses and number of flavors. For vanishing baryon chemical potential 
($\mu$) and for quark masses above a critical quark mass, a deconfinement transition associated 
with chiral restoration takes place through a smooth crossover~\cite{Aoki:2006we,Bazavov:2014pvz,Soltz:2015ula,Bazavov:2017xul}.
The study and characterization of the QGP gives information on the crossover transition 
as well as insights on the equation of state of deconfined matter~\cite{Andronic:2017pug,Bernhard:2016tnd}. 
These transitions are expected to have occurred in the early universe and therefore 
their study is also of relevance to cosmology~\cite{Boyanovsky:2006bf}.

Heavy-ion collisions at relativistic energies offer the possibility of studying the QGP 
by creating systems of dense matter at very high temperatures. Of the many observables 
that probe the QGP, measurements of \pai and \e meson production over a large transverse 
momentum (\pT) range and in different colliding systems are of particular interest. At 
low \pT (\pT~$<$~3~\GeVc), light meson production in heavy-ion collisions gives insights 
about hadronization and collectivity in the evolution of the QGP. At high-\pT 
(\pT~$>$~5~\GeVc), it helps quantify parton energy loss mechanisms~\cite{Burke:2013yra,Wang:2002ri}. 
High-\pT particle suppression in heavy-ion collisions with respect to pp collisions may 
be modified by cold nuclear matter effects, such as nuclear parton distribution function 
(nPDF) modifications with respect to the vacuum. Measurements in \pA collisions are thus 
needed to disentangle cold nuclear effects from the observed high-\pT particle suppression 
in \AACol collisions.

Other interesting probes of the QGP that can benefit from neutral meson measurements are 
studies of direct photon and heavy-flavor production measurements~\cite{Wilde:2012wc,Abelev:2014gla}. 
The \pai and \e mesons are the two most abundant sources of decay photons (and electrons); 
as a consequence, they generate the primary background for these rare probes. The first 
measurement of direct photons at the LHC \cite{Adam:2015lda} employed \mT-scaling and the 
K$^{0}_{\mbox{\small s}}$ reference measurement to estimate the \e contribution to decay 
photons. Forthcoming direct photon and heavy-flavor measurements at the LHC will be able 
to use the \e measurement directly. 

Measurements of pion spectra at RHIC~\cite{Adare:2013esx,Adler:2001yq} at low transverse 
momentum were observed to be well described by thermal models that assume a hydrodynamic 
expansion of a system in local equilibrium~\cite{Broniowski:2001we}. The comparison of 
these models to data suggested the presence of a thermalized system of quarks and gluons 
formed in the early stages of the collision. At LHC energies, the thermal models that describe 
the RHIC data also describe the ALICE charged pion spectrum~\cite{Abelev:2013vea} for 
\pT~$>$~0.5~\GeVc. Modern versions of these models fold in their calculations hydrodynamic 
expansion, which accounts for transverse flow effects, simultaneous chemical and thermal 
freeze-out and inclusion of high mass resonance decays from the PDG~\cite{Patrignani:2016xqp}. 
Among the many models that aim at explaining low-\pT particle production, the equilibrium 
and chemical non-equilibrium statistical hadronization models (EQ SHM and NEQ SHM, 
respectively) have had their validity tested against LHC data from \pT~$>$~0.1~\GeVc. 
The physics picture behind the NEQ SHM is a sudden hadronization of the QGP, that leads 
to the apperance of additional non-equlibrium chemical potentials for light and strange 
quarks. The low \pT pion enhancement predicted by the NEQ relative to the EQ SHM can be 
interpreted as the onset of pion condensation in ultra-relativistic heavy-ion collisions 
at the LHC energies~\cite{Begun:2013nga,Begun:2015ifa,Abuki:2008wm,Migdal:1979je,Zee:706825}. 
Both predictions can be further tested by measuring \pai and \e production at LHC energies. 

In the early RHIC program, a suppression of high-\pT \pai production was observed in 
heavy-ion collisions when compared to scaled pp data~\cite{PhysRevLett.109.152301}. This suppression 
was interpreted as a consequence of the energy loss of the scattered partons in the QGP 
generated in the collisions. From these observations, it was deduced that the dense QGP 
medium is opaque to energetic (hard) colored probes. Regarding high-\pT particle production 
at the LHC, it must be considered that the energy density of the plasma is higher than 
measured at RHIC. This increase in energy density leads to a larger energy loss of high-\pT 
partons with respect to those at both lower \pT ($<$~3~\GeVc) and lower energy~\cite{Aamodt:2010jd,Abelev:2014ypa}.
Moreover, it has been observed that baryons and strange mesons exhibit similar suppression 
as that of pions above 10~\GeVc. The measurement of another light meson, the \e meson, 
provides additional information about mechanisms of particle production and energy loss, 
while the measurement of both mesons at higher \pT will give insight about the \pT 
dependence of the suppression in this region. 

The suppression due to the QGP can also be studied with the \etapai ratio. In heavy-ion 
collisions, gluons are expected to experience larger energy loss in the medium than quarks, 
due to gluons having a larger vertex coupling factor. The energy reduction due to the 
presence of the medium (jet quenching effect)~\cite{Armesto:2011ht} may alter gluon and 
quark fragmentation differently with respect to what is observed in pp collisions. These 
differences between gluon and quark energy loss may introduce a modification in the 
suppression patterns observed for \pai and \e mesons, due to a larger gluon component in 
the \e meson (note that the \e meson, unlike \pai, has a two-gluon component)~\cite{Kroll:2002nt}. 
An intermediate \pT enhancement of the \etapai ratio in \AACol collisions relative to pp 
collisions would be an indication of the plasma induced color dependence suppression~\cite{Dai:2015dxa,Gyulassy:1990ye,Baier:1996sk}. 
The magnitude of this enhancement is sensitive to the initial values of the jet transport 
parameters and thus could be used to quantify the suppression.

In this paper, we present \pai and \e meson production measurements from the ALICE experiment 
in the \pT range 1~$<$~\pT~$<$~20~\GeVc in Pb-Pb collisions at center-of-mass energy \sNN 
in two centrality classes, 0--10\% and 20--50\%. The results are measured at midrapidity 
using two complementary detection methods: the photon conversion method (PCM) and use of 
the Electromagnetic Calorimeter (EMCal)~\cite{Cortese:2008zza}. The \pai results in the 
0--10\% centrality class have been combined with the previously published \pai result 
measured with the PHOS calorimeter~\cite{Abelev:2014ypa}. The new \pai measurement is 
updated with ten times more statistics than the previous ALICE measurement~\cite{Abelev:2014ypa}, 
and extends the \pT reach from 12~\GeVc to 20~\GeVc. The \e measurement is the first 
measurement of its kind at the LHC and has a wider \pT reach than what was previously 
measured at RHIC~\cite{Adler:2006hu}. \\

The paper is organized as follows: a brief description of the detectors used and of the 
data sample is given in Section~\ref{section:detectorsanddatasample}. The analysis 
procedure is described in Section~\ref{section:methods}. The results and the comparison 
to other experimental measurements and to theoretical predictions are presented in 
Section~\ref{section:resultslhc} and~\ref{section:resultsmodels}, respectively. 

\section{Detector description and data sample}\label{section:detectorsanddatasample}

The ALICE experiment and its performance are described in detail in \cite{ALICE,Abelev:2014ffa}.
The main detectors used for the reconstruction of \pai and \e mesons are located in the 
central barrel, operated inside a solenoidal magnetic field of 0.5~T directed along the 
beam axis. 

The Inner Tracking System (ITS) is a high granularity and precision detector that measures 
the position of the primary collision vertex and the impact parameter of the 
tracks~\cite{ALICE_ITS}. The ITS is composed of six cylindrical layers of silicon detectors 
positioned at radial distances from 4 to 43~cm. The two innermost layers of the ITS are 
Silicon Pixel Detectors (SPD) that cover the pseudorapidity regions $| \eta | <$~2 and 
$| \eta | <$~1.4. The next two layers are Silicon Drift Detectors (SDD) covering 
$| \eta | <$~1, while the two outer layers are Silicon Strip Detectors (SSD) covering 
$| \eta | <$~0.9. 

The Time Projection Chamber (TPC)~\cite{ALICE_TPC} is the main charged particle tracking 
and identification detector in the ALICE central barrel. It is a cylindrical drift detector 
filled with a Ne-CO$_{2}$ (90\%-10\%) gas mixture. This detector surrounds the ITS and 
is centered around the Interaction Point (IP) at a radial distance from 85 to 250~cm. The 
TPC has full azimuthal coverage and covers $| \eta | <$~0.9 for the full track length. 
Particles are identified through the measurement of their specific energy loss (\dEdx) 
in the detector with a 6.5\% resolution in the 0--5\% most central \Pb~\cite{Abelev:2014ffa,ALICE_TPC}. 
The track's transverse momentum resolution is $(\sigma(\pT)/\pT)$~=~0.8\% at 1~\GeVc 
and 1.7\% at 10~\GeVc in central \Pb collisions~\cite{Abelev:2014ffa,Abelev:2012hxa}.

The EMCal~\cite{Cortese:2008zza} is a sampling calorimeter composed of 77 alternating 
layers of 1.4~mm lead and 1.7~mm polystyrene scintillators. The EMCal is a fairly high 
granularity detector. It has a cell area of $\Delta\eta\times\Delta\phi$~=~0.0143~$\times$~0.0143~rad 
and an energy resolution of $\sigma_{E(\rm GeV)}/{E} = 4.8\%/E \oplus 11.3/\sqrt{E} \oplus 1.7\%$~\cite{Allen20106}. 
In year 2011, it covered $| \eta | <$~0.7 and $\Delta \varphi$~=~100~degrees. 

The main detectors used for triggering and characterization of the collision are the 
V0~\cite{Abbas:2013taa} and the Zero Degree Calorimeters (ZDC)~\cite{ALICE:2012aa}. The 
V0 consists of two scintillator arrays located on opposite sides of the Interaction Point 
(IP) at 340 and 90~cm covering 2.8~$< \eta <$~5.1 and $-$~3.7~$ < \eta < -$~1.7, respectively. 
The ZDC detectors are located at a distance of 114~m on both sides of the IP and detect spectator nucleons.

The \Pb data sample used for this analysis was collected in the 2011 LHC run. During that 
period, about 358 ion bunches circulated in each LHC beam, with collisions delivering a 
peak luminosity of 4.6~$\times$~10$^{-4} \mu$b$^{-1}$s$^{-1}$, corresponding to an 
average of about 10$^{-3}$ hadronic interactions per bunch crossing. The minimum bias (MB) 
trigger was defined by the coincidence of signals in the two V0 arrays synchronized with 
a bunch crossing. An online selection based on the measured V0 amplitudes was employed to 
enhance central (0--10\%) and semi-central (0--50\%) events~\cite{Abelev:2014ffa}. The 
ZDC and the V0 were also used for the rejection of pile-up and beam-gas interactions. The 
centrality class definition was based on the V0 amplitude distributions. The number of 
binary collisions (N$_{\rm coll}$) for a given value of the centrality was extracted with 
the help of a Glauber model~\cite{Glauber} as detailed in \cite{ Abelev:2012hxa,Abelev:2013qoq}. 
Only events with a reconstructed primary vertex within $| z_{\mbox{\tiny vtx}} | <$~10~cm 
of the nominal interaction vertex along the beam direction were accepted. The data are 
analyzed in two centrality classes: 0--10\% and 20--50\%, containing 1.9~(1.6)~$\times$~10$^{7}$ 
and 1.3~(1.1)~$\times$~10$^{7}$ events for PCM (EMCal), respectively. The minimum bias 
trigger cross section, $\sigma^{\mbox{\tiny PbPb}}_{\mbox{\tiny MB}}$~=~(7.64$\pm$0.22(syst.))~b~\cite{Abelev:2013qoq},
was determined using van der Meer scans~\cite{vanderMeer:296752}. The integrated luminosity, 
corresponding to the number of analyzed events normalized by $\sigma^{\mbox{\tiny PbPb}}_{\mbox{\tiny MB}}$ 
in each centrality percentile, is 20.1~$\mu$b$^{-1}$ and 4.8~$\mu$b$^{-1}$ for 0--10\% and 
for 20--50\%, respectively. 

\section{Analysis methods}\label{section:methods}

The \pai and \e mesons are reconstructed using the two-photon decay channel, 
\pai$\rightarrow \gamma\gamma$ and \e$\rightarrow \gamma\gamma$, with a branching ratio 
of (98.823~$\pm$~0.034)\% and (39.41~$\pm$~0.20)\%~\cite{Patrignani:2016xqp}, respectively. 
With the photon conversion method, photons that convert in the detector material are 
measured by reconstructing the electron-positron pairs in the central rapidity detectors 
using a secondary vertex (V$^{0}$) finding algorithm~\cite{Abelev:2014ffa}. This method 
produces a V$^{0}$ candidate sample on which the analysis quality selection criteria were 
applied, as done in \cite{Abelev:2012cn,Abelev:2014ypa}. Electrons, positrons and photons 
are required to have $| \eta | <$~0.9. To ensure track quality, a minimum track momentum 
of 50~\MeVc and a fraction of TPC clusters over findable clusters (the number of 
geometrically possible clusters which can be assigned to a track) above 0.6 have been 
required. Moreover, a maximum conversion radius of 180~cm delimits the TPC fiducial volume 
for good track reconstruction, while a minimum of 5~cm rejects Dalitz decays of the type 
\pai(\e)$\rightarrow e^{+}e^{-}\gamma$. The specific energy loss \dEdx should be within 
the interval [$-$3~$\sigma_{\tiny \mbox{d}E/\mbox{d}x}$, $+$5$\sigma_{\tiny \mbox{d}E/\mbox{d}x}$] 
from the expected electron Bethe-Bloch parametrization value, where $\sigma$ is the 
standard deviation of the energy loss measurement. Pions are rejected by a selection of 
3$\sigma$ above the pion hypothesis in the range 0.4~$< p <$~2~\GeVc and of 1$\sigma$ for 
$p >$~2~\GeVc. The smaller rejection with respect to the previous \Pb measurement translates 
into a larger efficiency at high-\pT for the \pai and \e mesons. To further reject 
K$^{0}_{s}$, \xspace$\Lambda$ and \xspace$\overline{\Lambda}$ from the V$^{0}$ candidates, 
a selection is applied on the components of the momenta relative to the V$^{0}$, using 
the asymmetry of the longitudinal momentum of the V$^{0}$ daughters 
($\alpha_{V^{0}} = (p_L^{e^+} - p_L^{e^-})/(p_L^{e^+} + p_L^{e^-})$), and on the transverse 
momentum of the electron with respect to the V$^{0}$ momentum 
($q_{\mbox{\tiny T}} = p_{e} \times \sin{\theta_{\mbox{\tiny V$^{0}, e$}}}$). V$^{0}$ 
candidates are selected with a two-dimensional elliptic selection criterion of 
$( \alpha_{V^{0}} / \alpha_{V^{0}_{\mbox{\tiny max}}} )^{2} + ( q_{\mbox{\tiny T}} / q_{\mbox{\tiny T, max}} )^{2} <$~1, 
with $\alpha_{V^{0}_{\mbox{\tiny max}}}$~=~0.95 and $q_{\mbox{\tiny T, max}}$~=~0.05~\GeVc, 
in order to increase the purity while optimizing efficiency of the photon sample. As 
conversion electrons have a preferred decay orientation, a selection on $\psi_{\mbox{\tiny pair}}$, 
the angle between the plane perpendicular to the magnetic field and the plane containing 
the electron and positron tracks, together with a cut on the photon $\chi^{2}$ of the 
Kalman filter~\cite{KalmanFilter}, further suppresses the contamination from non-photonic 
V$^{0}$ candidates. This cut, described in \cite{Acharya:2017tlv}, is applied requiring 
$\chi^{2}_{\gamma, \rm max}$~=~20 and $\psi_{\rm pair,max}$~=~0.1. To improve the signal 
significance, a \pT-dependent cut on the energy asymmetry of the photons 
$|\alpha| <$~0.65$\cdot \tanh({1.8 \mbox{(\GeVc)$^{-1}$} \cdot\it{p}_\text{T}})$ 
(where $\alpha = (E_{\gamma_1} - E_{\gamma_2})/(E_{\gamma_1} + E_{\gamma_2})$, \pT in 
\GeVc) is applied.

For the measurement with the EMCal, photons stemming from meson decays are measured directly. 
Photon-like hits in the detector are identified by energy deposits in the neighboring cells, 
which are grouped into clusters with a minimum size of 2 cells. A minimum energy per cell 
of 50~MeV is required. The cluster finding algorithm employs a seed energy of 
$E_{\mbox{\tiny seed}}$~=~0.3~GeV, which is slightly above the minimum ionizing particle 
threshold~\cite{Abelev:2014ffa}. EMCal clusters that coincide within a window of $|\Delta \eta|<$~0.025 
and $|\Delta \phi| <$~0.05 radians of a charged particle reconstructed in the TPC and projected 
to the EMCal surface are rejected. Each selected EMCal cluster is then required to have a 
total energy of at least 1.5~GeV to remove low energy pairs consisting of predominantly 
combinatorial background and particle conversions in the material. A loose photon-like 
electromagnetic shower shape selection is applied to the clusters by looking at the eccentricity 
of the cluster via the weighted RMS of the shower energy along the major ellipse axis according to
\begin{equation}
\sigma_{\rm long}^{2} = \frac{s_{\eta\eta}+s_{\varphi\varphi} }{2} +\sqrt{\frac{\left(s_{\eta\eta}-s_{\varphi\varphi}\right)^{2}}{4}+s^{2}_{\eta\varphi}},
\end{equation}
where $s_{ij} = \langle ij \rangle - \langle i \rangle\langle j \rangle$ are the covariance 
matrix elements, $i,j$ are cell indices in $\eta$ or $\varphi$ axes,$\langle ij \rangle$ 
and $\langle i \rangle$, $\langle j \rangle$ are the second and the first moments of the 
cluster cells weighted with the cell energy logarithm~\cite{Grassmann:1984px,Alessandro:2006yt,Abelev:2014ffa,Acharya:2017hyu}. 
The purpose of this loose shower shape selection 0.1~$< \sigma_{long}^{2} <$~0.5 (photons 
sit in a narrow peak centered at 0.25) is to remove noisy and very deformed or asymmetric 
cluster shapes which result from the merging of different particle showers produced nearby 
in the calorimeter.

\begin{figure}[!h]
\centering
\includegraphics[width=0.4\textwidth]{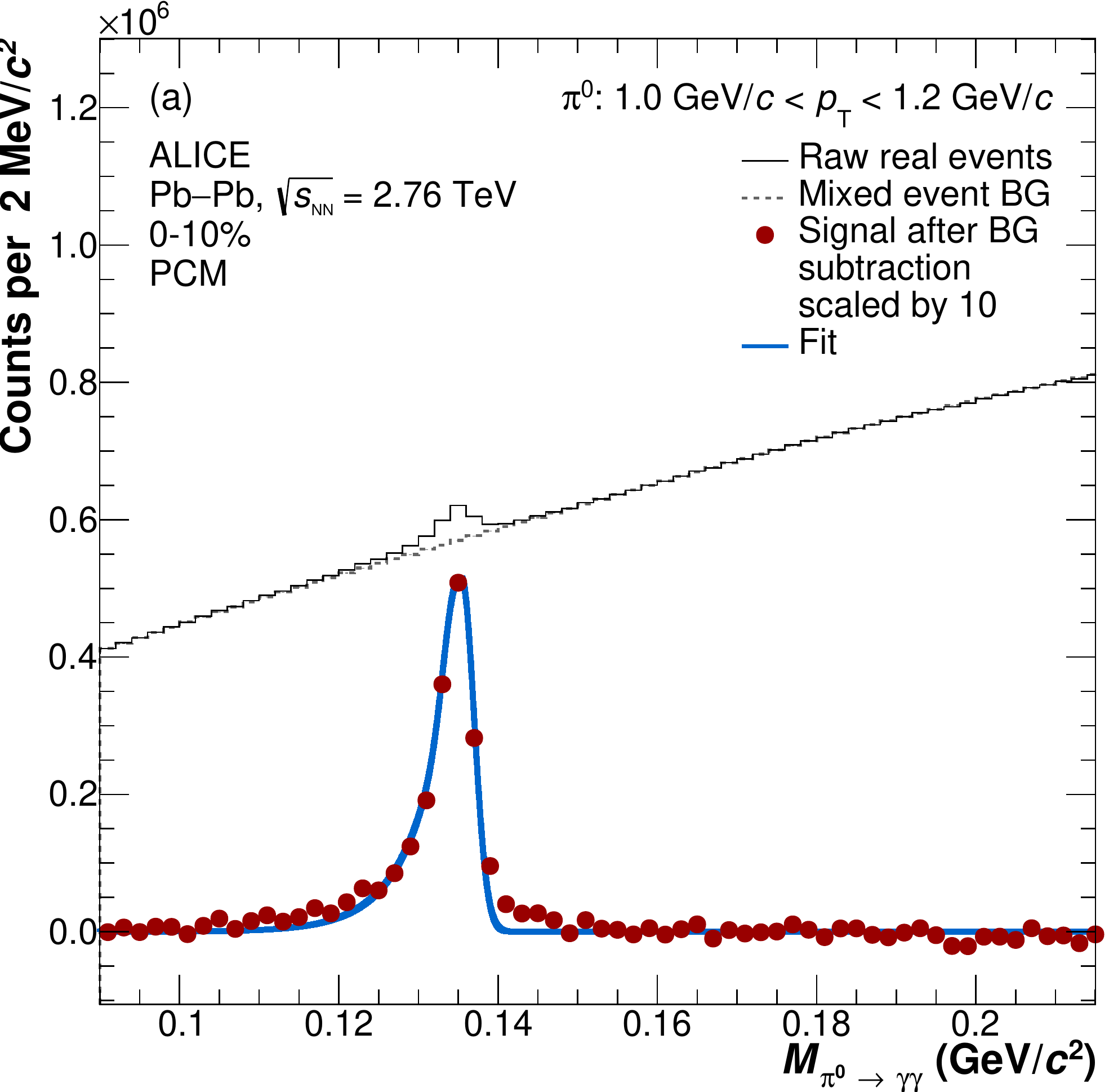}\hspace*{0.3cm}
\includegraphics[width=0.4\textwidth]{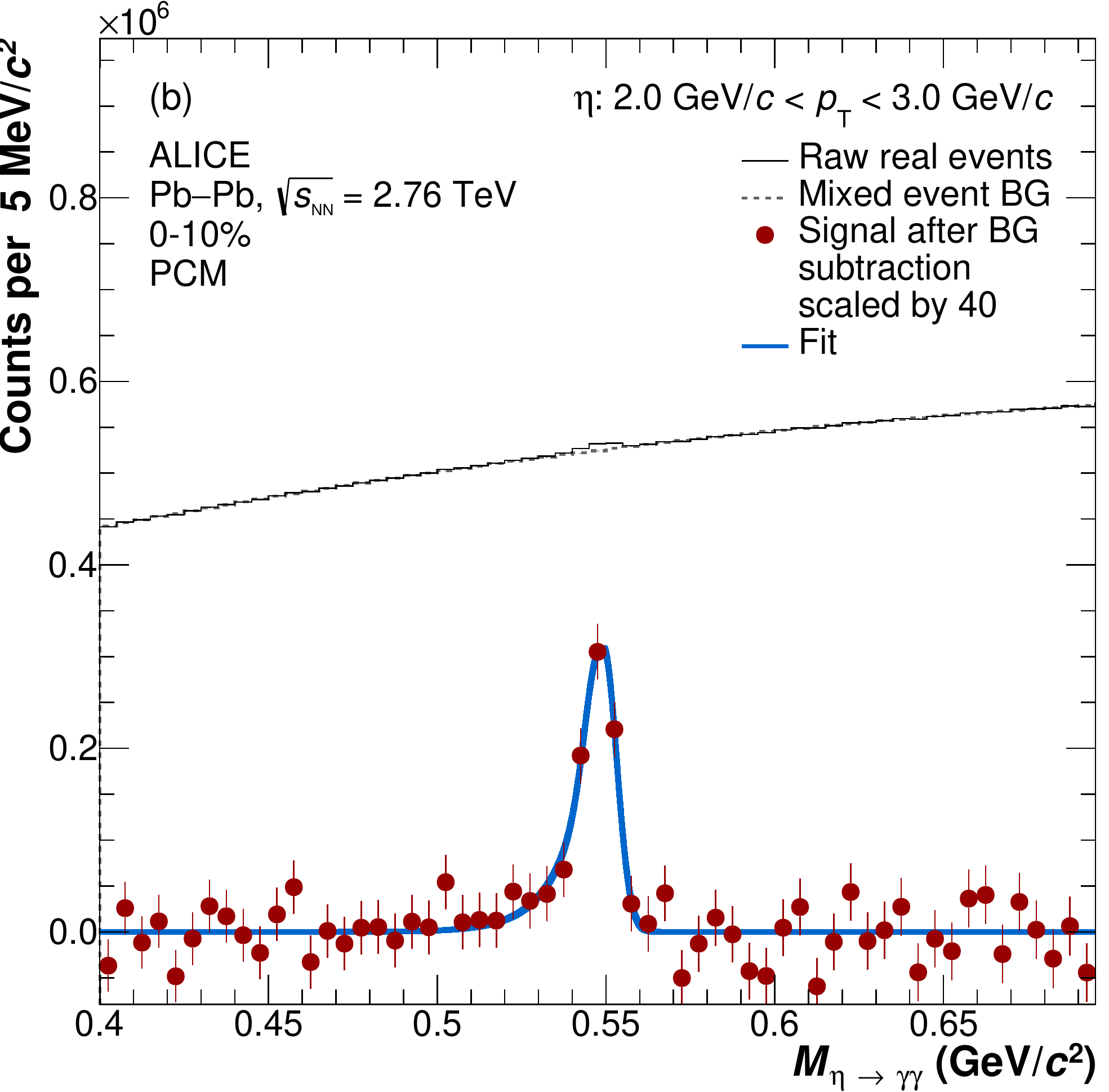}\\
\includegraphics[width=0.4\textwidth]{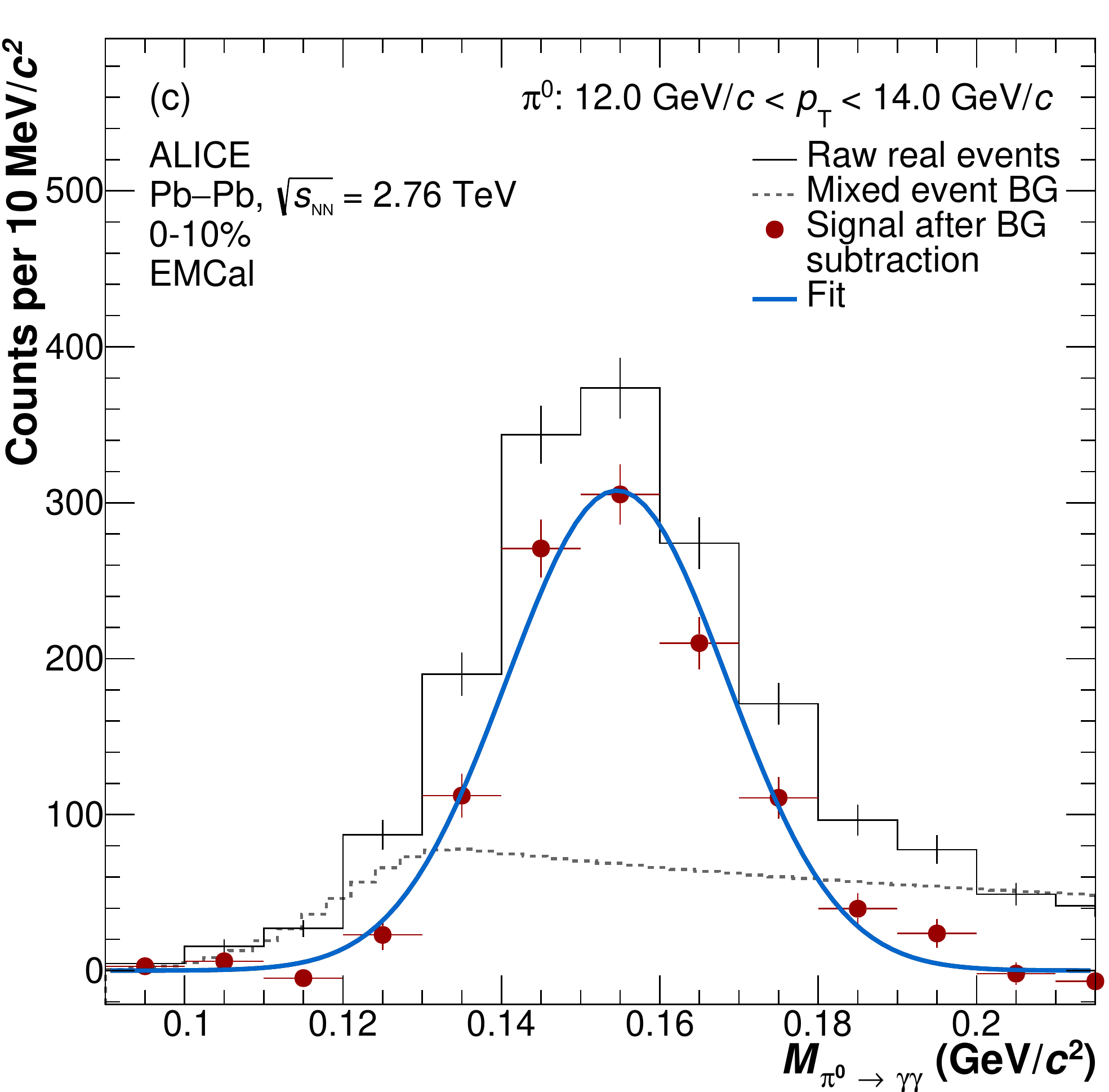}\hspace*{0.3cm}
\includegraphics[width=0.4\textwidth]{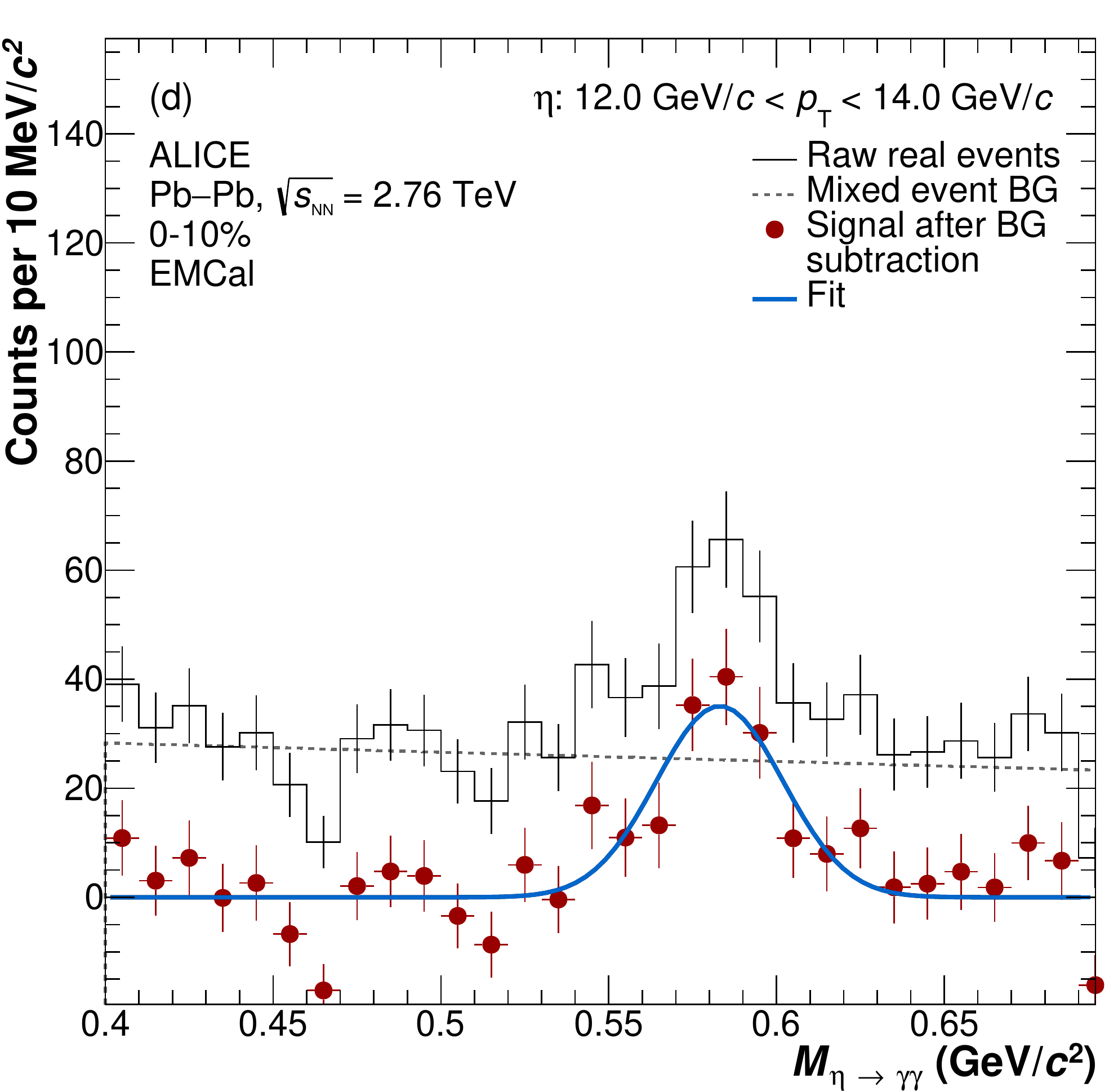}
\caption{(Color online) Invariant mass distribution of reconstructed photon pairs $M_{\gamma\gamma}$ 
for the \pai and \e mesons measured with PCM, (a) and (b), and EMCal, (c) and (d), in the 
centrality class 0--10\%. The black histograms show the signal before background subtraction 
while the red bullets show the signal after subtraction. The estimated background is indicated 
by the grey dashed lines. The blue lines are the fit to the invariant mass peak after the 
combinatorial and residual background subtraction (see text for description).}
\label{fig:InvMassPi0}
\end{figure}

For the PCM and EMCal analyses, the reconstructed two-photon invariant mass is measured in 
bins of \pT in the rapidity range $| y | <$~0.85 and $| y | <$~0.7, respectively. The \pT 
ranges in which the separate methods contribute are reported in Table~\ref{ranges}. In 
addition, a minimum photon pair opening angle of 5~mrad is used to reject background in 
the PCM analysis. 
\begin{table}[!h]
\vspace*{0.2cm}
\centering
\begin{tabular}{r|c|c|c|c|c|}  
\cline{2-6}
		&\multicolumn{3}{| c |}{\paitable}		&\multicolumn{2}{ c |}{\e } \\ 
\cline{2-6}
		&PCM		&EMCal		&PHOS  		&PCM		&EMCal 	 \\ 
\hline
0--10\%		&1~--~14~\GeVc 	&4~--~20~\GeVc	&1~--~12~\GeVc 	&1~--~10~\GeVc	&4~--~20~\GeVc \\
20--50\%	&1~--~14~\GeVc 	&4~--~20~\GeVc	& --		&1~--~10~\GeVc	&4~--~20~\GeVc \\
\hline
\end{tabular}
\caption{Transverse momentum ranges for the \pai and \e meson measurements. For the \e meson 
in both centralities and for the \pai in 20--50\% centrality class the combination is between 
PCM and EMCal. For \pai in the 0--10\%, the final results are obtained combining PCM, EMCal 
as well as previously published results using the PHOton Spectrometer (PHOS)~\cite{Abelev:2014ypa}.}
\label{ranges}
\end{table}\\
The background under the neutral meson signal contains combinatorial and correlated contributions. 
The combinatorial background is estimated with the event mixing method by mixing photons from 
different events but with similar photon multiplicity and topological (vertex location on the $z$ 
axis, and in the particular case of the PCM analysis the event plane angle) characteristics. 
The mixed event background is normalized to the reconstructed two-photon invariant mass in a 
region at higher mass with respect to the meson peak and subtracted. Additionally, various 
fitting functions for the total background are also used in order to obtain the number of 
mesons and to evaluate the corresponding systematic uncertainty (EMCal). The resulting 
invariant mass distributions are fit with either a Gaussian combined with a low mass exponential 
tail~\cite{Matulewicz1990194} (PCM, to account for electron bremsstrahlung) on top of a linear 
function (PCM, to account for residual background) or with a Crystal Ball distribution~\cite{Gaiser:1985ix} 
(EMCal) in order to obtain the position and width of the peak~\cite{Abelev:2014ffa}. After 
subtracting the total background, the yields are extracted for each \pT bin by integrating 
the invariant mass distributions over a range that depends on the peak position and resolution. 
Fig.~\hyperref[fig:InvMassPi0]{\ref*{fig:InvMassPi0}} shows the invariant mass distribution for 
the \pai and \e mesons reconstructed with PCM and EMCal. 
\\
Corrections for geometrical acceptance, reconstruction efficiency, secondary \pai from weak 
decays (the measured spectra of the relevant particles~\cite{Abelev:2013xaa} are taken as 
input) and hadronic interactions and occupancy effects due to cluster overlaps (for EMCal) 
were estimated with a Monte Carlo simulation using HIJING~\cite{Gyulassy:1994ew} as the event 
generator. The simulated particles are propagated through the apparatus via GEANT3~\cite{GEANT3}, 
where a realistic detector response based on experimental conditions is applied in order 
to reproduce the performance of the ALICE detector during data taking. The simulated events 
are then analyzed with the same reconstruction and analysis selection criteria applied to 
the experimental data. 

It was verified that the detector resolutions were well reproduced by the Monte Carlo 
simulations~\cite{Abelev:2014ffa}. The mass peak positions and widths measured in the data 
for each centrality interval for the PCM (EMCal) analysis were reproduced within 0.5$\%$ 
(1.5$\%$) or better, and the remaining discrepancies have been taken into account in the 
systematic uncertainties associated with the difference of the energy scale and position 
of the calorimeter between data and Monte Carlo.

In the PCM analysis, the pile-up contribution is estimated by analyzing the distance of 
closest approach distribution for the photon candidates, as done in \cite{Abelev:2014ypa}. 
The effect of pile-up in the EMCal analysis was verified to be negligible since the EMCal 
cell timing resolution is an order of magnitude better than the bunch crossing spacing of 
200~ns used in the 2011 \Pb run.

For both methods, the systematic uncertainties were studied by varying the selection 
criteria used in the two analyses and by studying the resulting variations of the fully 
corrected spectra in individual \pT bins. The largest contribution to the systematic 
uncertainties for the PCM analysis comes from the uncertainty in the material budget~\cite{Abelev:2014ffa}, 
and amounts to 9\%. Other sources of systematic uncertainties include the yield extraction, 
track reconstruction, electron identification and photon reconstruction (mainly for the 
\e meson). The details of the PCM systematic uncertainties are listed in Table~\ref{tab:SysErrs}.

The main source of systematic uncertainties for the neutral meson detection with the EMCal 
is associated with the particle identification criteria used to select photon pairs (PID). 

The uncertainties due to the signal extraction in a given $p_{T}$ interval are taken as 
the mean of the uncertainties obtained in all signal and background parametrizations. 
Variations on the values used for the meson identification selection criteria are also 
included and the root mean square (RMS) of these values is used as a systematic uncertainty.

The EMCal detector energy response was determined by analyzing test beam data~\cite{Allen20106}. 
Comparisons of the mass peak position and the energy-to-momentum ratios of electron 
tracks~\cite{Adam:2016khe} in data and Monte Carlo simulations quantify the overall 
systematic uncertainty due to the Monte Carlo description of the energy response and 
position of the calorimeter. This uncertainty amounts to 8.6\% of the invariant yield 
measurements. 

Other sources of systematic uncertainties are the material budget, the \pT distribution 
of the simulations used for the extraction of efficiencies and the contribution from 
higher mass decays. The details of the EMCal systematic uncertainties are listed in 
Table~\ref{tab:SysErrs}.
\begin{table*}[tbh]
\centering
\vspace*{0.2cm}
  \resizebox{\linewidth }{!}{ %
\begin{tabular}{c|c|c|c|c|c|c|c|c|} 
\cline{2-9}
					    &\multicolumn{8}{c|}{PCM}   \\
\cline{2-9}
					    &\multicolumn{4}{c|}{0--10\%}			   			&\multicolumn{4}{c|}{20--50\%}   \\
\cline{2-9}
					    &\multicolumn{2}{c|}{\paitable} 	&\multicolumn{2}{c|}{$\eta$}	 	&\multicolumn{2}{c|}{\paitable} 	&\multicolumn{2}{c|}{$\eta$} \\
\cline{2-9}
					    &1.1~GeV/$c$  &5.5~GeV/$c$		&2.5~GeV/$c$  	&5.0~GeV/$c$		&1.1~GeV/$c$  	&5.5~GeV/$c$		&2.5~GeV/$c$	&5.0~GeV/$c$ \\
\hline
Material budget                             & 9.0   	  & 9.0     		& 9.0		& 9.0  			& 9.0		& 9.0			& 9.0	& 9.0  	\\
Track reconstruction 		      	    & 2.3  	  & 2.6    		& 6.0		& 6.2			& 1.4		& 2.3	   		& 7.0   & 9.0 	 	 \\
Yield extraction                            & 1.5      	  & 2.1    		& 6.4		& 7.0  			& 2.5		& 2.8	  		& 10.0	& 11.0 		\\
$e^{+}/e^{-}$ identification                & 1.7      	  & 2.5    		& 6.0	  	& 6.1 			& 1.4		& 2.4			& 5.5	& 9.3		 \\
Photon reconstruction			    & 3.7  	  & 2.1    		& 13.7	 	& 13.6	 		& 2.1		& 2.2			& 8.0 	& 8.6 \\
\hline
\hline
					    &  \multicolumn{8}{c|}{EMCal}   \\
\cline{2-9}
					    &\multicolumn{4}{c|}{0--10\%}			   			&\multicolumn{4}{c|}{20--50\%}   \\
\cline{2-9}
					    &\multicolumn{2}{c|}{\paitable} 	&\multicolumn{2}{c|}{$\eta$}	 	&\multicolumn{2}{c|}{\paitable} 	&\multicolumn{2}{c|}{$\eta$} \\
\cline{2-9}
					    &7.0~GeV/$c$	& 18.5~GeV/$c$	& 7.0~GeV/$c$	& 18.5~GeV/$c$  	&7.0~GeV/$c$	&18.5~GeV/$c$	& 7.0~GeV/$c$	& 18.5~GeV/$c$ \\
\hline
Signal extraction        		    &2.9		&5.1		&4.2		&5.5			&7.5		&5.8		&6.0		&7.1\\
Photon identification    		    &9.5		&8.0		&4.6		&6.0			&7.5		&4.5		&14.1		&5.0\\
Energy response             		    &8.6		&8.6		&8.6		&8.6			&8.6		&8.6		&8.6		&8.6\\
Material budget          	  	    &5.0		&5.0		&5.0		&5.0			&5.0		&5.0		&5.0		&5.0\\
Hijing simulation        		    &8.6		&10.0		&8.6		&10.0			&2.0		&5.3		&2.0		&5.3\\
Monte Carlo input        	            &2.0		&3.0		&$<$1		&1.5			&$<$1		&$<$1		&$<$1		&$<$1\\
Higher mass decays       		    &4.0		&2.0		&-   		&-  			&3.2		&2.0		&-		&-\\
\hline
\hline
\end{tabular}}
\caption{Summary of the systematic uncertainties in percent for selected \pT regions for 
the PCM and EMCal analyses.}
\label{tab:SysErrs}
\end{table*}\\
When computing the \etapai ratio and the nuclear modification factor, fully and partially 
correlated errors, such as material budget and energy scale (EMCal only), are taken into 
account.

\section{Results}\label{section:resultslhc}

\subsection{Invariant yields of the $\pi^{0}$ and $\eta$ meson}
\begin{figure}[!h]
\centering
\includegraphics[width=0.495\textwidth]{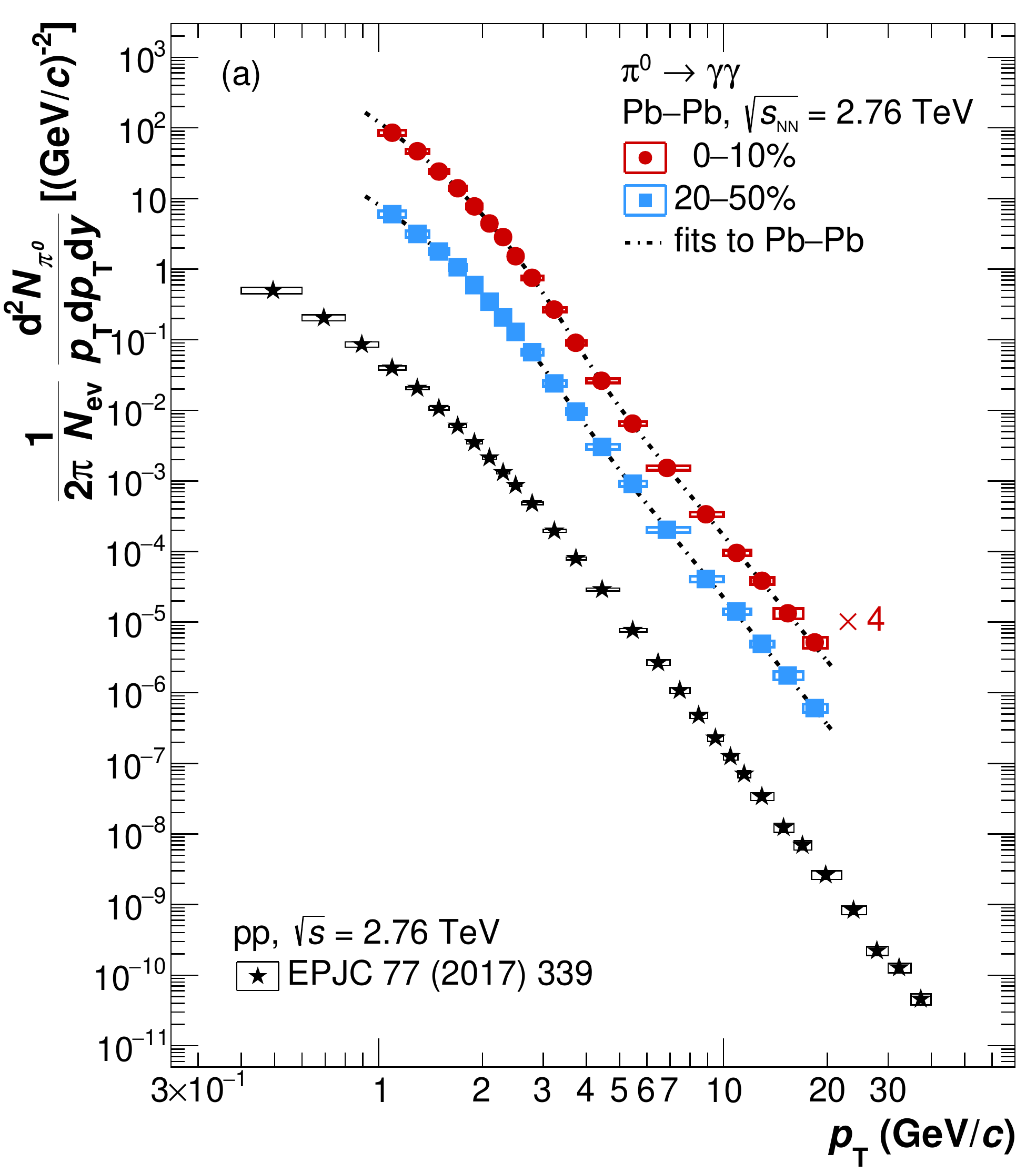} 
\includegraphics[width=0.495\textwidth]{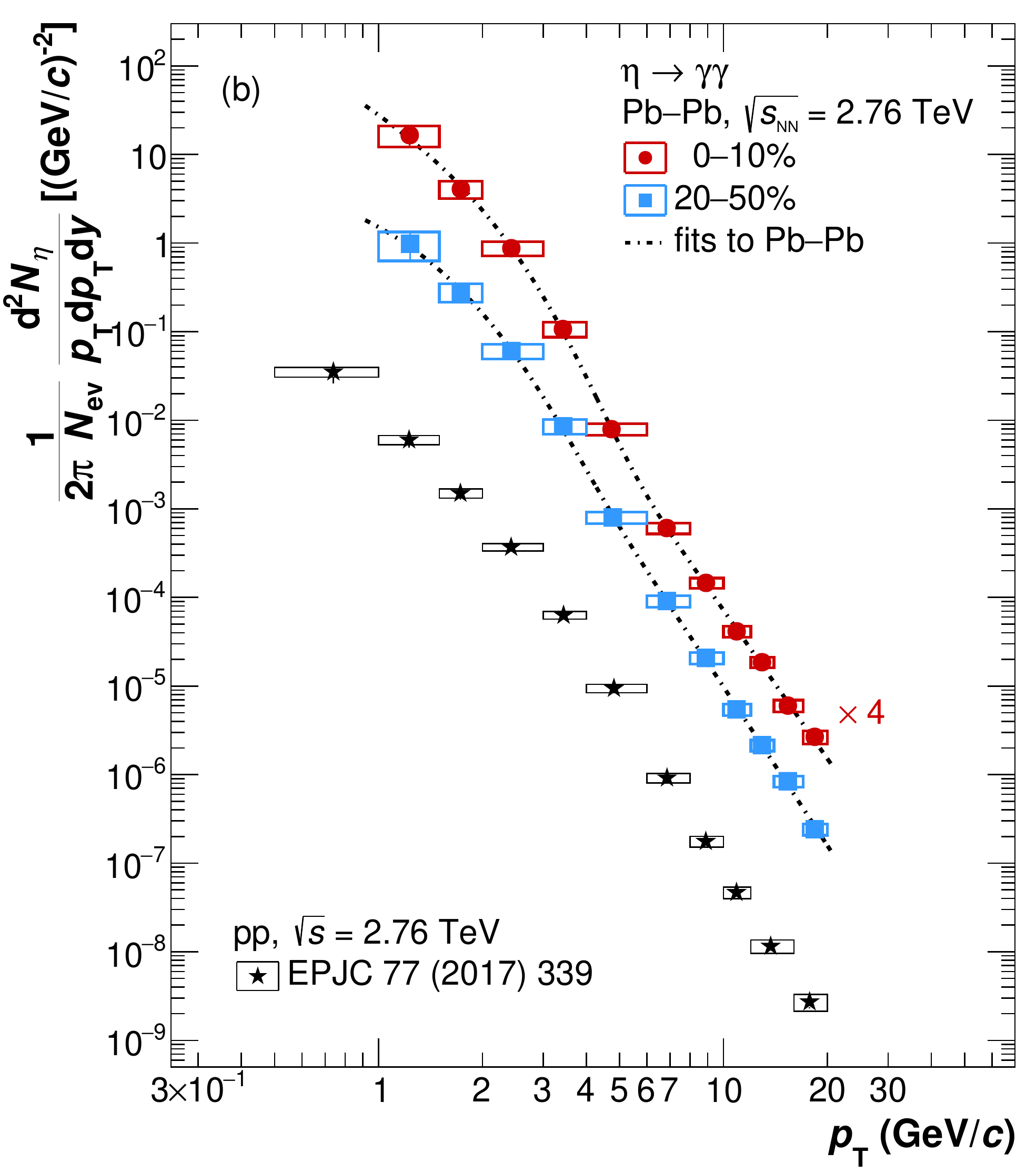}
\caption{(Color online) Invariant yields of the (a) \pai and (b) \e meson in the centrality 
classes 0--10\% (circles) and 20--50\% (squares). The vertical error bars represent the 
statistical uncertainties while the boxes represent the systematic uncertainties. The \Pb 
measurements are compared with the corresponding pp invariant cross sections (stars) measured 
at the same center-of-mass energy~\cite{Abelev:2014ypa,Acharya:2017hyu}. The dashed black lines 
correspond to the fits to the data with the two-component function. See Table~\ref{FitParam} 
and corresponding text for details.}
\label{fig:InvYieldsDataOnly}
\end{figure}

The invariant differential yields for \pai and \e mesons have been calculated employing
\begin{equation}\label{eq:yields}
E \frac{\mbox{d}^3 N}{\mbox{d}p^3}  = \frac{1}{2\pi N_{\mbox{\tiny evt}}}\frac{1}{B_{\mbox{\tiny Ratio}}~ A\varepsilon } \frac{N_{\mbox{\tiny raw}}}{ p_{\mbox{\tiny T}} \Delta p_{\mbox{\tiny T}} \Delta y},
\end{equation}
where $N_{\mbox{\tiny evt}}$ is the number of events in the centrality class considered, 
$B_{\rm Ratio}$ is the branching ratio~\cite{Patrignani:2016xqp} for the process 
$\pi^{0}(\eta) \rightarrow \gamma\gamma$, A$\varepsilon$ are the corresponding acceptance 
and efficiency corrections and $N_{\mbox{\tiny raw}}$ corresponds to the reconstructed 
$\pi^{0}(\eta)$ raw yield within the rapidity range $\Delta y$ and the transverse momentum 
bin $\Delta p_{\mbox{\tiny T}}$. The horizontal location of the data points is shifted 
towards lower \pT from the bin center by a few MeV and illustrates the \pT value where 
the differential cross section is equal to the measured integral of the cross section over 
the corresponding bin~\cite{Lafferty:1994cj}. For the \e/\pai ratio and \RAA the bin-shift 
correction is done in y-coordinates. The \pT ranges in which the measurements were performed 
are reported in Table~\ref{ranges}. In the overlap region a weighted average of the two 
results (or three when applicable) is performed using the inverse of the quadratic sum of 
the uncertainties (statistical and systematic) that are uncorrelated between the methods 
as weights~\cite{Lyons:1988rp,Valassi:2003mu,Valassi:2013bga}. \\
Fig.~\ref{fig:InvYieldsDataOnly} shows the invariant differential yields of (a) \pai and 
(b) \e meson measured in pp~\cite{Acharya:2017hyu} and \Pb collisions in the two centrality 
bins under study. The \pai meson measurements are in agreement with the previously published 
ALICE \pai spectra~\cite{Abelev:2014ypa} and extend the transverse momentum reach from 12 to 
20~\GeVc. For the \e meson, the results presented here are the first measurement of its 
kind in heavy-ion collisions at the LHC and the first measurement of this meson to reach 
down to \pT of 1~\GeVc in a collider experiment~\cite{Adler:2006hu,Adler:2006bv}. 

Both meson spectra have been parametrized over the full \pT range by the function proposed 
in \cite{Bylinkin1,Bylinkin2} that combines a Boltzmann factor at low-\pT with a power law 
at high-\pT
\begin{equation}\label{eq:Bylinkin}
E \frac{\mbox{d}^3 N}{\mbox{d}p^3} = A_{e} \exp{\frac{- \left( \sqrt{p_{\mbox{\tiny T}}^{2} + M^{2}} - M \right)}{T_{e}}} + \frac{A}{ \left(1 + \frac{p_{\mbox{\tiny T}}^{2}}{T^{2}n} \right)^{n}}
\end{equation}
where $M$ is the meson mass (in \massGeVc), $A_{e}$, $A$, $T_{e}$, $T$ and $n$ are free 
parameters of the fit. The parameters resulting from the fits to the meson invariant yields 
in both centrality classes are reported in Table~\ref{FitParam}. All parameters are free 
except for the amplitude $A$. The values are chosen after a systematic study of the two 
separate components of the Bylinkin--Rostovtsev function and of the parameter limits variation.
\begin{table}[!h]
\centering
\begin{tabular}{l|c|c|c|c|}  
\cline{2-5}
&\multicolumn{2}{c |}{\pai}  &\multicolumn{2}{c |}{\e } \\ 
\cline{2-5}
		& 0--10\% 			& 20--50\% 			& 0--10\% 			& 20--50\% \\ 
\hline
$A_{e}$ (GeV/c)$^{-2}$	&162~$\pm$~20	&30~$\pm$~7 			&15~$\pm$~6			&4.2~$\pm$~2.5 \\
$T_{e}$ (\GeVc) &0.37~$\pm$~0.01		&0.38~$\pm$~0.02		&0.44~$\pm$~0.03		&0.42~$\pm$~0.06 \\
$A$ (GeV/c)$^{-2}$		&840 		&80 				&100	 			&2  \\
$T$ (\GeVc)	&0.34~$\pm$~0.01		&0.50~$\pm$~0.02		&0.38~$\pm$~0.03		&0.76~$\pm$~0.05\\
$n$ 		&3.00~$\pm$~0.05		&3.00~$\pm$~0.05		&3.0~$\pm$~0.1			&3.0~$\pm$~0.1 \\
$\chi^{2}/ndf$	&0.18				&0.20				&0.22				&0.14		\\
\hline
\end{tabular}
\caption{Parameters of the fits to the differential invariant yields of \pai and \e meson 
using the two-component function of A.A.~Bylinkin and A.A.~Rostovtsev~\cite{Bylinkin1, Bylinkin2}. 
The total uncertainties, i.e. quadratic sum of statistical and systematic uncertainties, 
are used for the fits.}
\label{FitParam}
\end{table}

\subsection{Particle ratios}
\begin{figure}[h]
\centering
\includegraphics[width=0.495\textwidth]{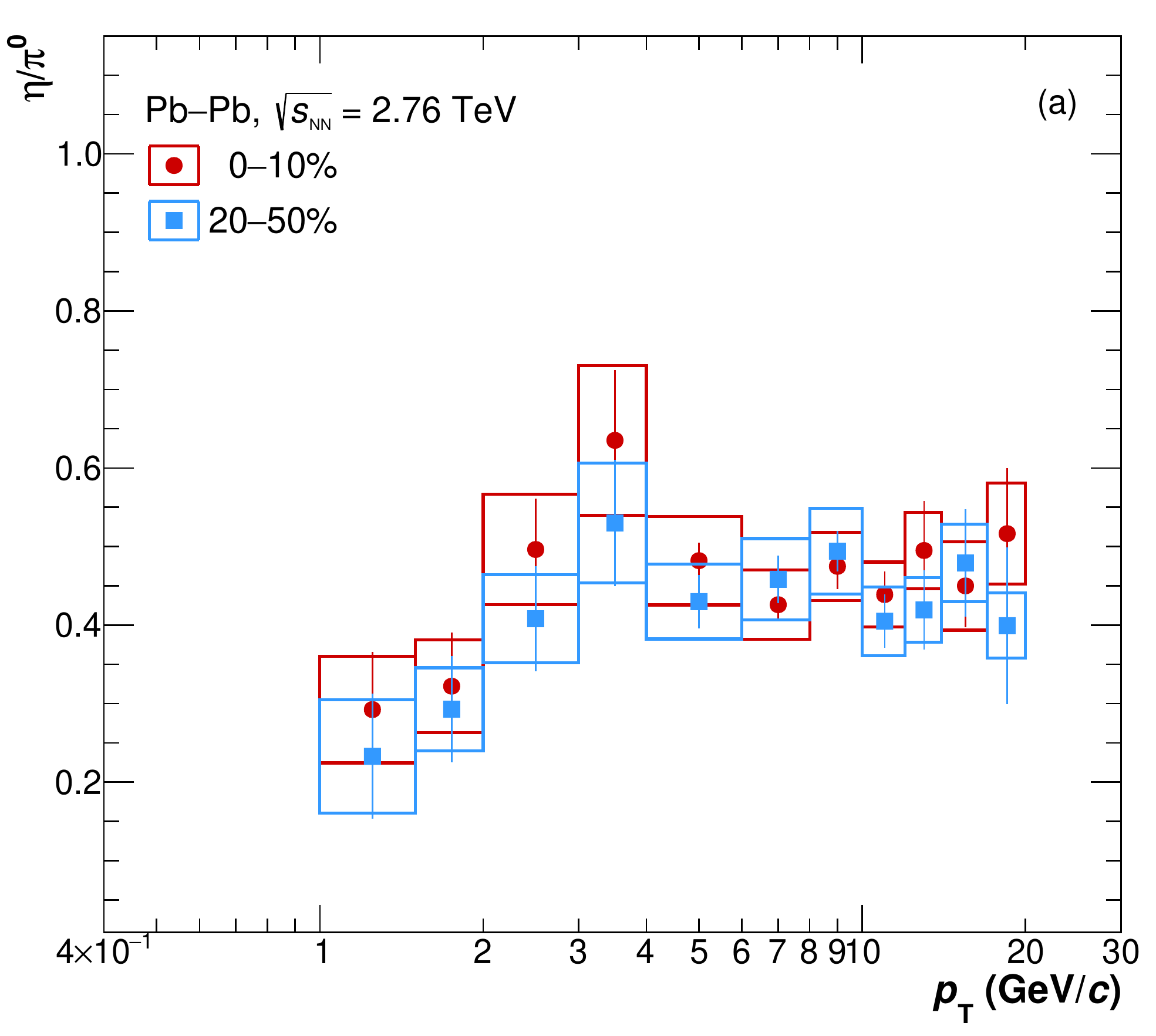}
\includegraphics[width=0.495\textwidth]{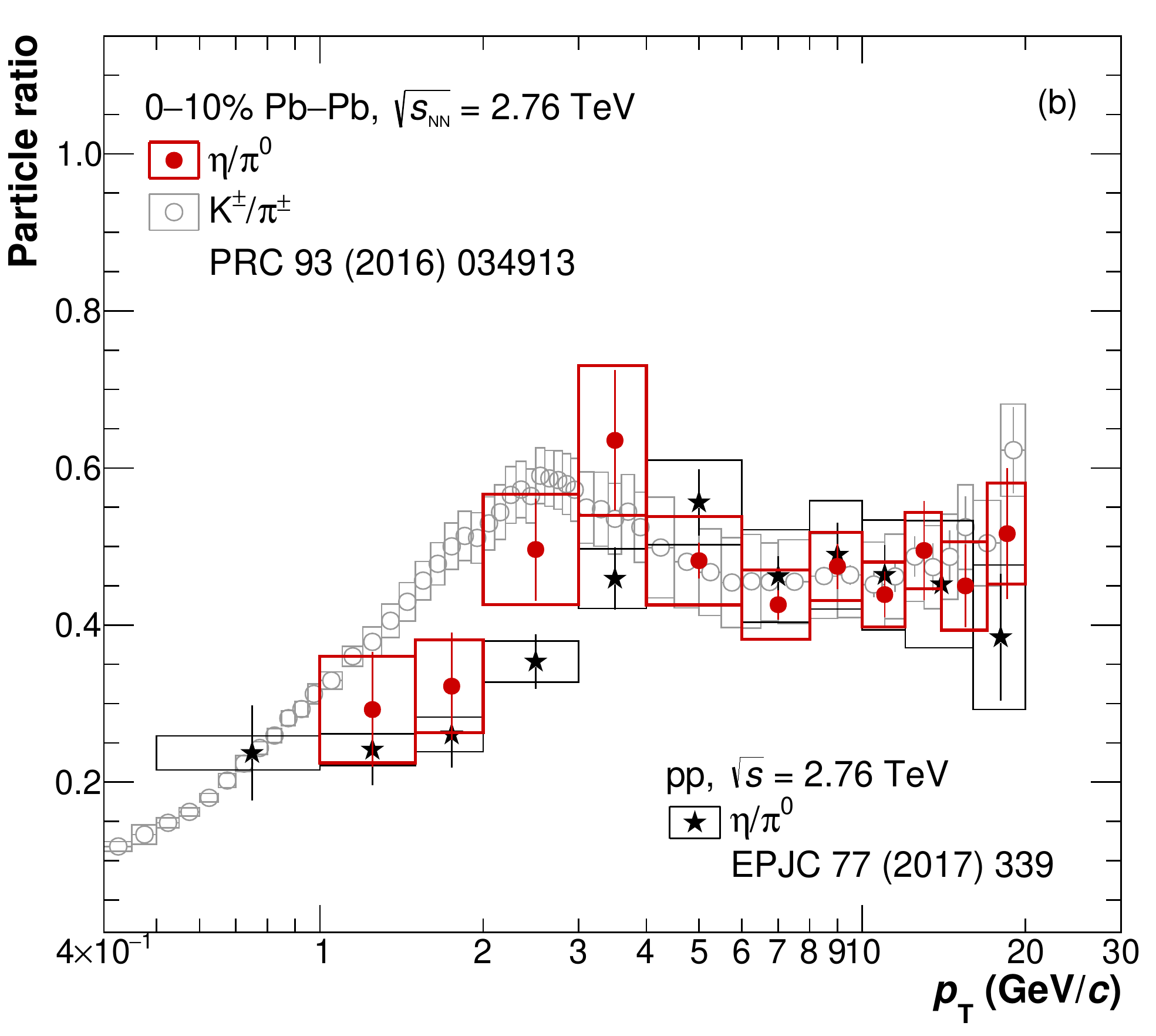}
\caption{(Color online) (a) \etapai ratio in the two centrality classes measured, 0--10\% 
(circles) and 20--50\% (squares). (b) Comparison of the \etapai measurement in the 0--10\% 
centrality class (full circles) to the corresponding ratio in pp collisions~\cite{Acharya:2017hyu} 
(stars) and to the $K^{\pm}/\pi^{\pm}$ measurement in the same centrality class, system 
and collision energy~\cite{Abelev:2013vea} (open circles).}
\label{fig:RatioEtaToPi}
\end{figure}

The \etapai ratio measured in the two centrality classes is shown in Fig.~\ref{fig:RatioEtaToPi}~(a). 
In Fig.~\ref{fig:RatioEtaToPi}~(b), the measurement in the 0--10\% centrality class is 
compared to the same ratio measured in pp collisions at $\sqrt{s}$~=~2.76~TeV~\cite{Acharya:2017hyu}, 
as well as to the $K^{\pm}/\pi^{\pm}$ ratio in the same centrality class and in the same 
collision system and energy~\cite{Abelev:2013vea}, measured by ALICE. The $K^{\pm}/\pi^{\pm}$ 
ratio is of interest as the relative mass differences between these particles is similar 
to the one for the \e and \pai mesons. At \pT~$<$~2~\GeVc, the \etapai and the $K^{\pm}/\pi^{\pm}$ 
ratios in Pb--Pb are in agreement within uncertainties. At 2~$<$~\pT~$<$~4~\GeVc, due to 
the large uncertainties in the \etapai ratio in \Pb, no conclusion can be made on the 
significance of the difference between the \etapai ratio in pp or the $K^{\pm}/\pi^{\pm}$ 
ratio in \Pb. At \pT~$>$~4~\GeVc, the value for all ratios is of similar magnitude. 
Moreover, a constant fit from 3 to 20~\GeVc gives a plateau value for the ratio of 
0.457~$\pm$~0.013$^{stat}$~$\pm$~0.018$^{syst}$, in agreement with the value quoted in 
lower center-of-mass energy measurements~\cite{Adler:2006hu}.

\subsection{The nuclear modification factor \RAA}
\begin{figure}[h]
\centering
\includegraphics[width=0.495\textwidth]{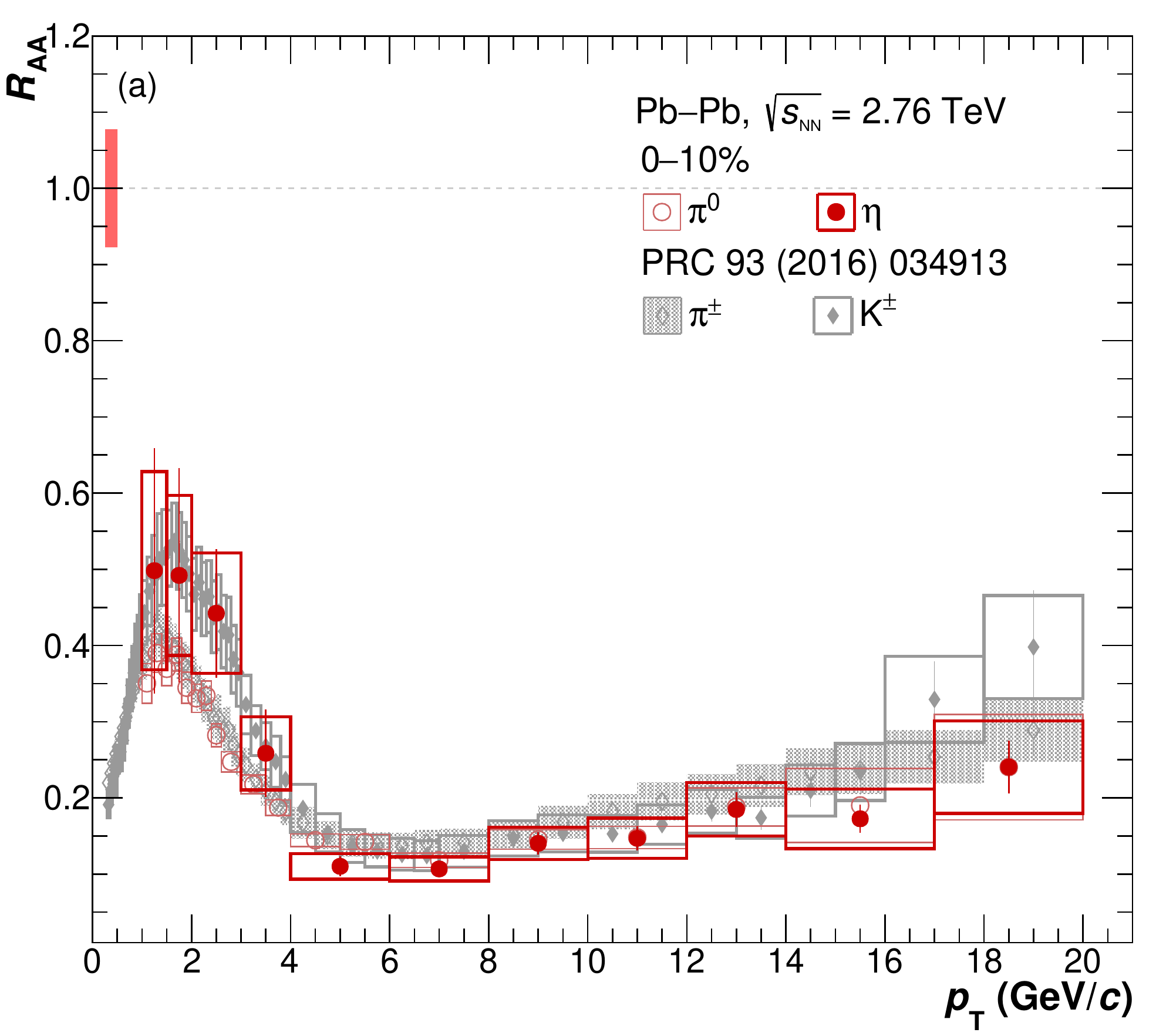}
\includegraphics[width=0.495\textwidth]{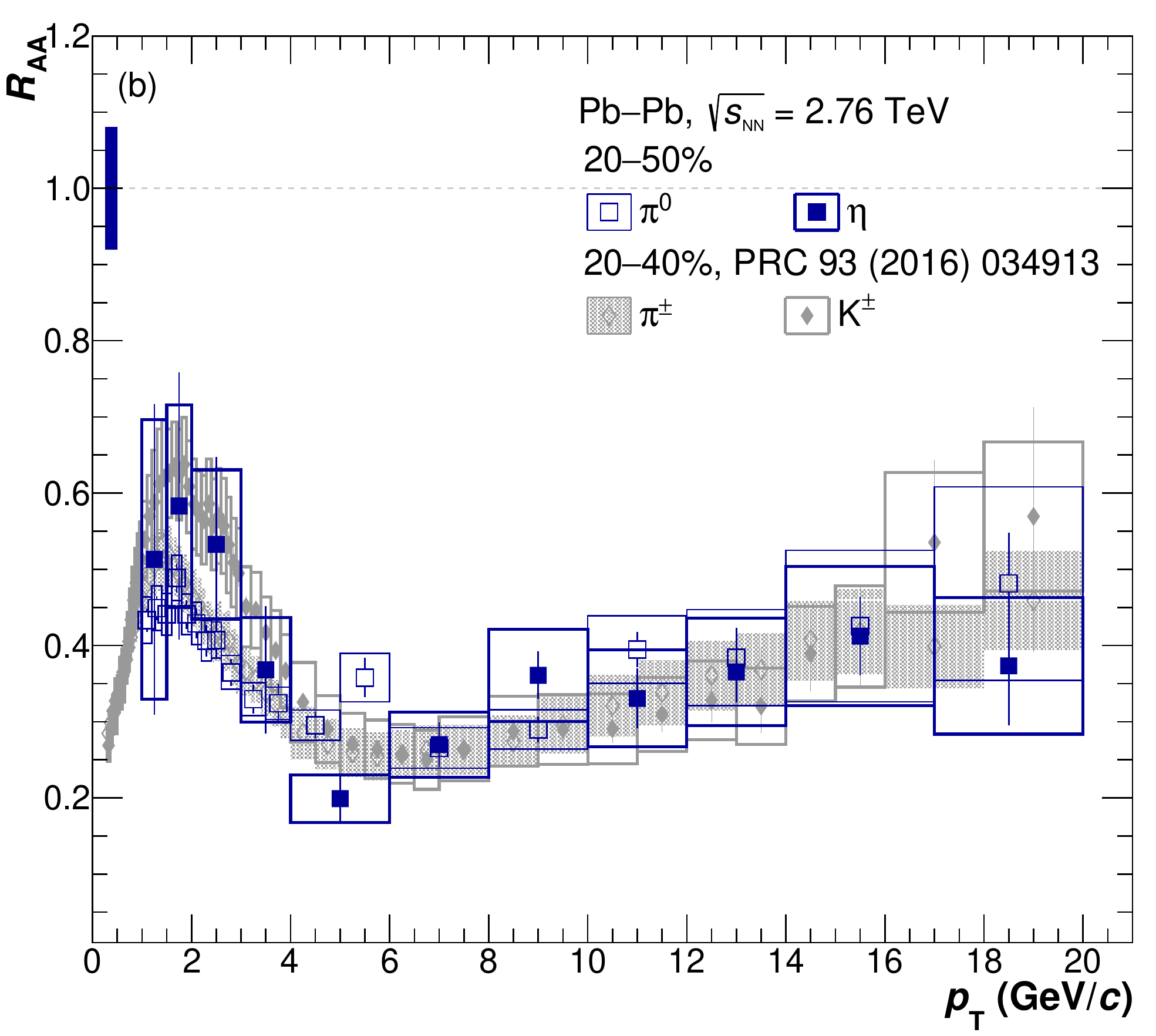} 
\caption{(Color online) Measured nuclear modification factor for the \pai (empty symbols) 
and \e meson (full symbols) in the (a) 0--10\% and (b) 20--50\% centrality classes, compared 
to ALICE $\pi^{\pm}$ and $K^{\pm}$~\cite{Adam:2015kca, Abelev:2014laa} (open and full diamonds) in the same 
centrality classes. The boxes around unity represent quadratic sum of the uncertainty on 
$\langle \TAA \rangle$ and on the pp spectrum normalization uncertainty.}
\label{fig:RAABothMesons}
\end{figure}

The nuclear modification factor can be used to quantify particle production suppression 
in heavy-ion collisions with respect to pp collisions. It is defined as
\begin{equation}
\RAA (\pT) = \frac{ \mathrm{d}^{2}N/\mathrm{d}\pT \mathrm{d}y|_\mathrm{AA}}{\langle \TAA \rangle \times
\mathrm{d}^{2}\sigma/\mathrm{d}\pT \mathrm{d}y|_\mathrm{pp} }.
\end{equation}
where the nuclear overlap function $\langle \TAA \rangle$ is related to the average 
number of inelastic collisions by $\langle \TAA \rangle =\langle N_{\rm{coll}} \rangle/\sigma_{\rm{inel}}^{\rm{pp}}$ 
and $\sigma_{\rm{inel}}^{\rm{pp}}$ is the total inelastic cross-section determined using 
van der Meer scans~\cite{Abelev:2012sea}. \\
The mean number of collisions is 1501~$\pm$~165 for the centrality class 0--10\% and 
349~$\pm$~34 for the centrality class 20--50\%~\cite{Abelev:2013qoq}. The \pai and \e 
meson spectra measured in pp collisions at the same center-of-mass energy are obtained 
from \cite{Acharya:2017hyu}.\\
The measured \RAA is presented in Fig.~\ref{fig:RAABothMesons} for the \pai and the \e 
mesons. A \pT and centrality dependent suppression is clearly observed. For the most 
central collisions, the \RAA has a maximum around \pT~$\approx$~1.5~\GeVc and a minimum 
for \pT~$\approx$~7~\GeVc, after which it increases. The increase at high-\pT could be 
due to the variation of the relative gluon and quark contributions to meson production 
as a function of \pT, with gluons being expected to suffer a stronger suppression than 
quarks due to a larger Casimir factor~\cite{Sarkar:2010zza}. \\
The suppression observed at high-\pT is consistent with recent ATLAS results~\cite{Aaboud:2017bzv}, 
and may indicate a larger quark than gluon relative contribution for high-\pT jet 
production in heavy-ion collisions at the LHC. A similar behavior is observed for 
semi-central events, though with a smaller suppression over the full transverse momentum 
range. The magnitude and pattern of the suppression is the same for the \pai and \e 
mesons for \pT~$>$~4~\GeVc despite the difference in mass. At lower \pT, the present 
accuracy is not enough to determine if the suppression is different for the two mesons.
The \RAA values for both centrality classes are also compared to the ALICE charged kaon 
\RAA~\cite{Adam:2015kca} measured at the same center-of-mass energy and collision system 
(Fig.~\ref{fig:RAABothMesons}), and is of interest given the similar masses of kaons and \e 
mesons. This comparison indicates similar suppression patterns for $\eta$ and $K^{\pm}$ 
across the whole \pT range and similar suppression between all particles for \pT~$>$~4~\GeVc. 
This result is consistent with previous baryon and strange meson \RAA results~\cite{Adam:2015kca, Abelev:2014laa} 
indicating that the energy loss in the medium is likely a purely partonic effect. 

\subsection{Comparisons to lower energy measurements}
\begin{figure}[h]
\centering
\includegraphics[width=0.495\textwidth]{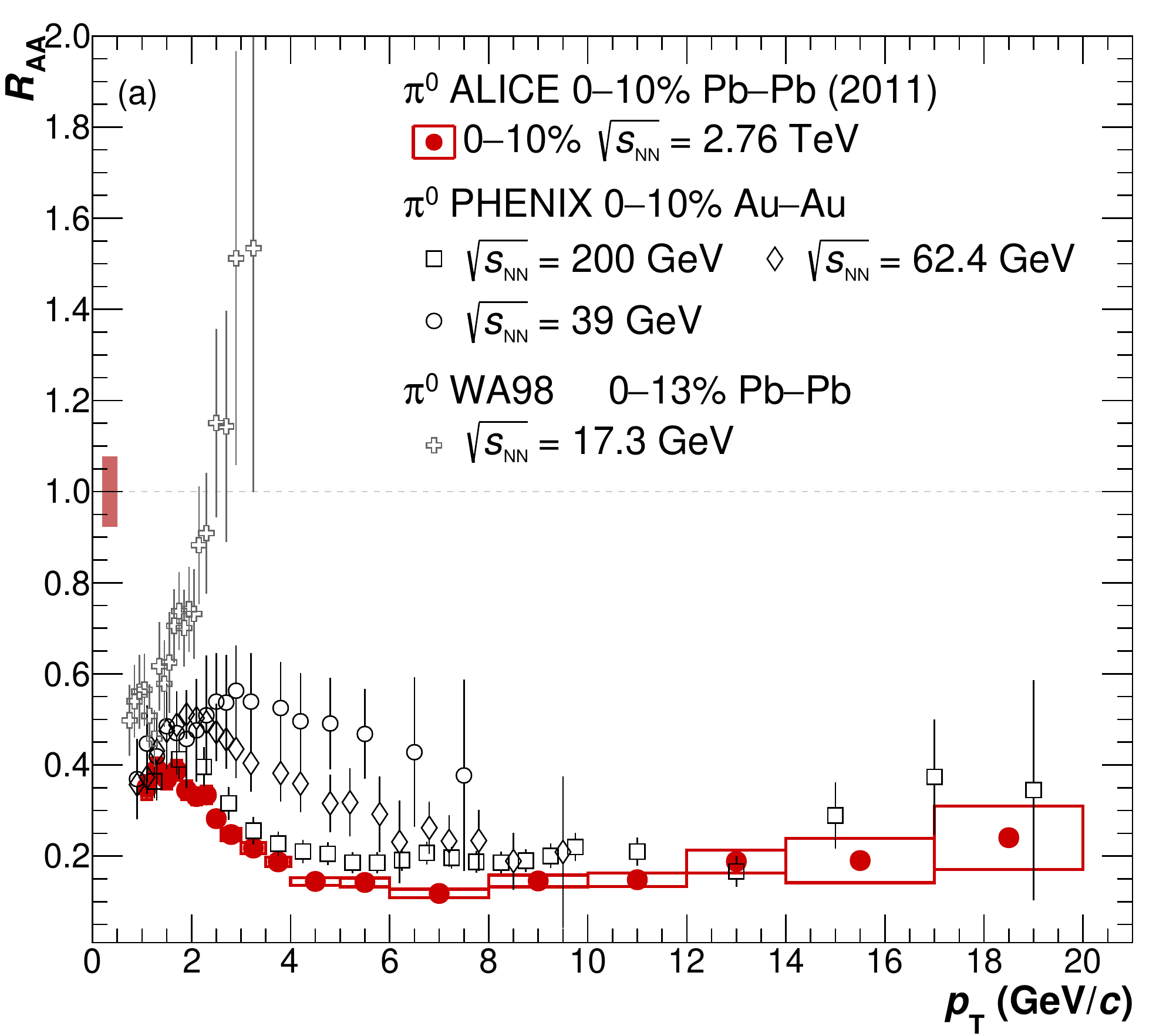}
\includegraphics[width=0.495\textwidth]{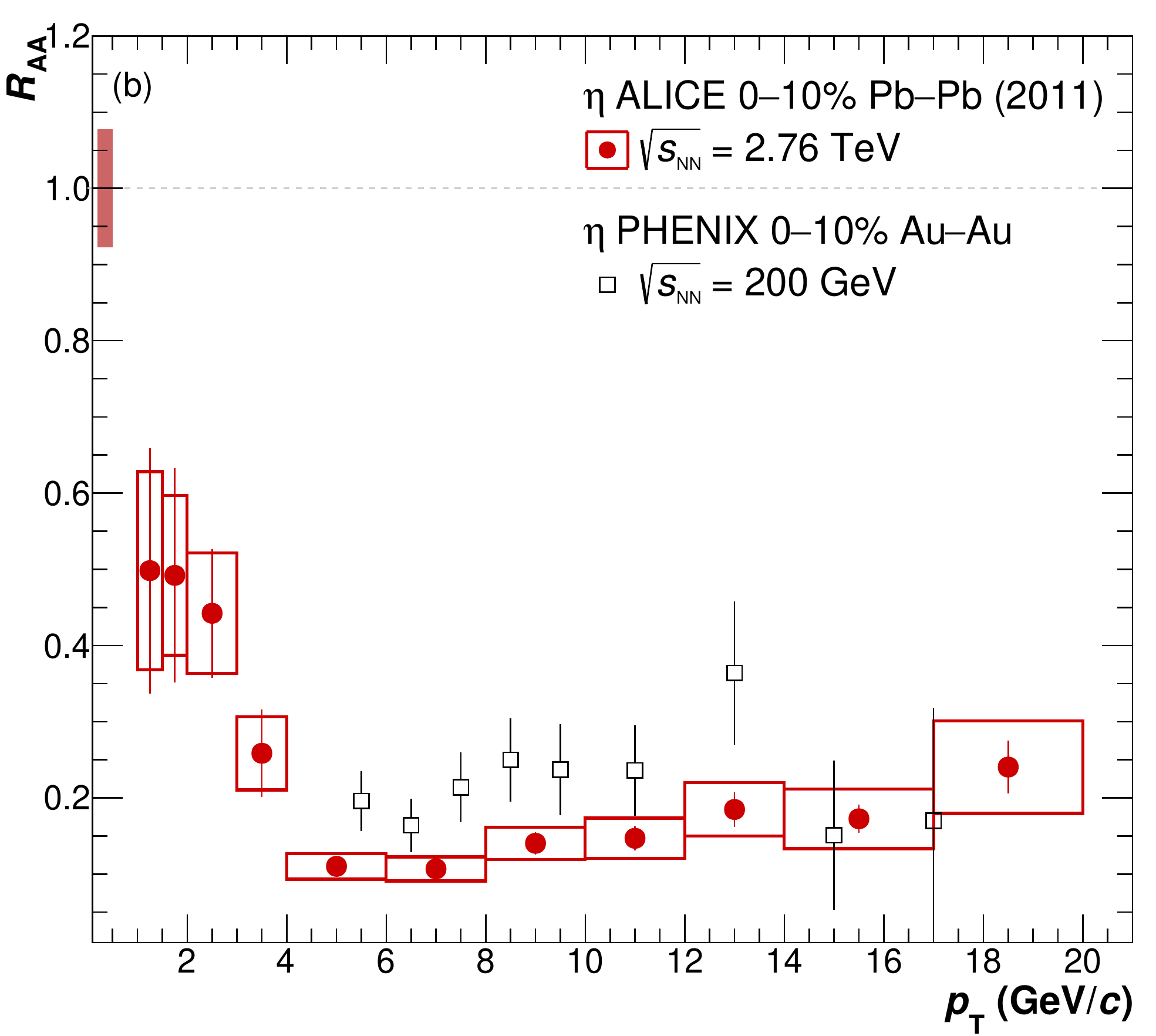}
\caption{(Color online) \RAA of the (a) \pai and (b) \e meson compared to data from lower 
center-of-mass energy results~\cite{Adare:2008qa,PhysRevLett.109.152301,Adler:2006hu}.}
\label{fig:RAAWorldDataPion}
\end{figure}

The nuclear modification factor in the 0--10\% centrality class is compared to previous \pai 
measurements reported by the WA98~\cite{PhysRevLett.100.242301} and PHENIX 
collaborations~\cite{Adare:2008qa,PhysRevLett.109.152301} (Fig.~\ref{fig:RAAWorldDataPion}, 
(a)) for center-of-mass energies per binary collision $\sqrt{s_{\mbox{\tiny NN}}}$ ranging 
from $17.3$~GeV (WA89) to $200$~GeV (PHENIX).\\
Our results confirm a dependence of the suppression on the center-of-mass energy and indicate 
a larger suppression for increasing collision energy. At \pT~$>$~11~\GeVc, the relative difference 
in suppression between the PHENIX and ALICE data is inconclusive due to the large uncertainties.  \\
The \e meson \RAA is compared to the corresponding PHENIX measurement~\cite{Adler:2006hu} at 
\sNNR (Fig.~\ref{fig:RAAWorldDataPion}, (b)). Similarly to the \pai case, the ALICE measurement 
shows a larger suppression compared to the PHENIX data in the region 5~$<$~\pT~$<$~14~\GeVc.

\section{Comparisons to models}\label{section:resultsmodels}
\begin{figure}[!h]
\centering
\includegraphics[width=0.54\textwidth,valign=c]{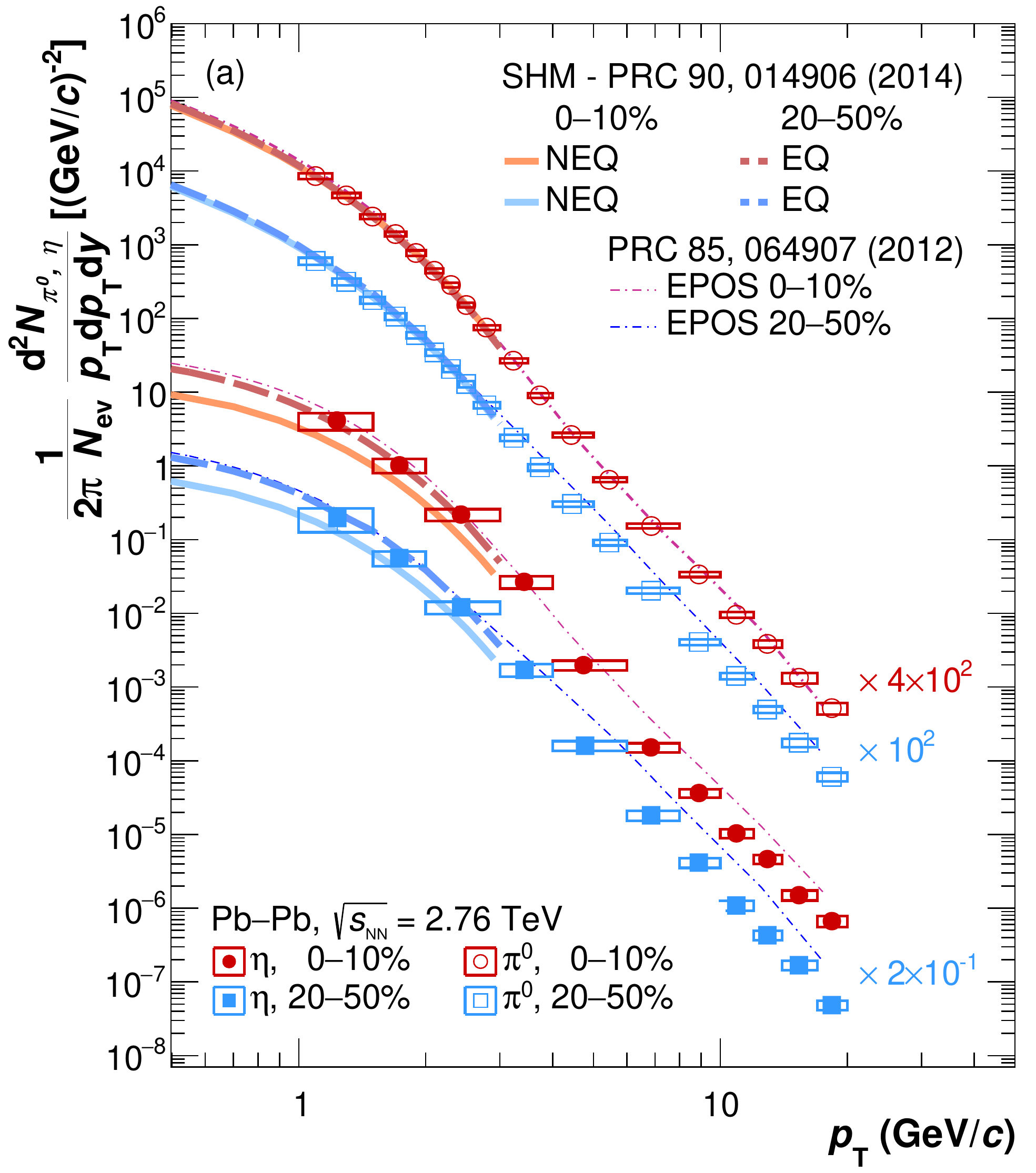}
\includegraphics[width=0.45\textwidth,valign=c]{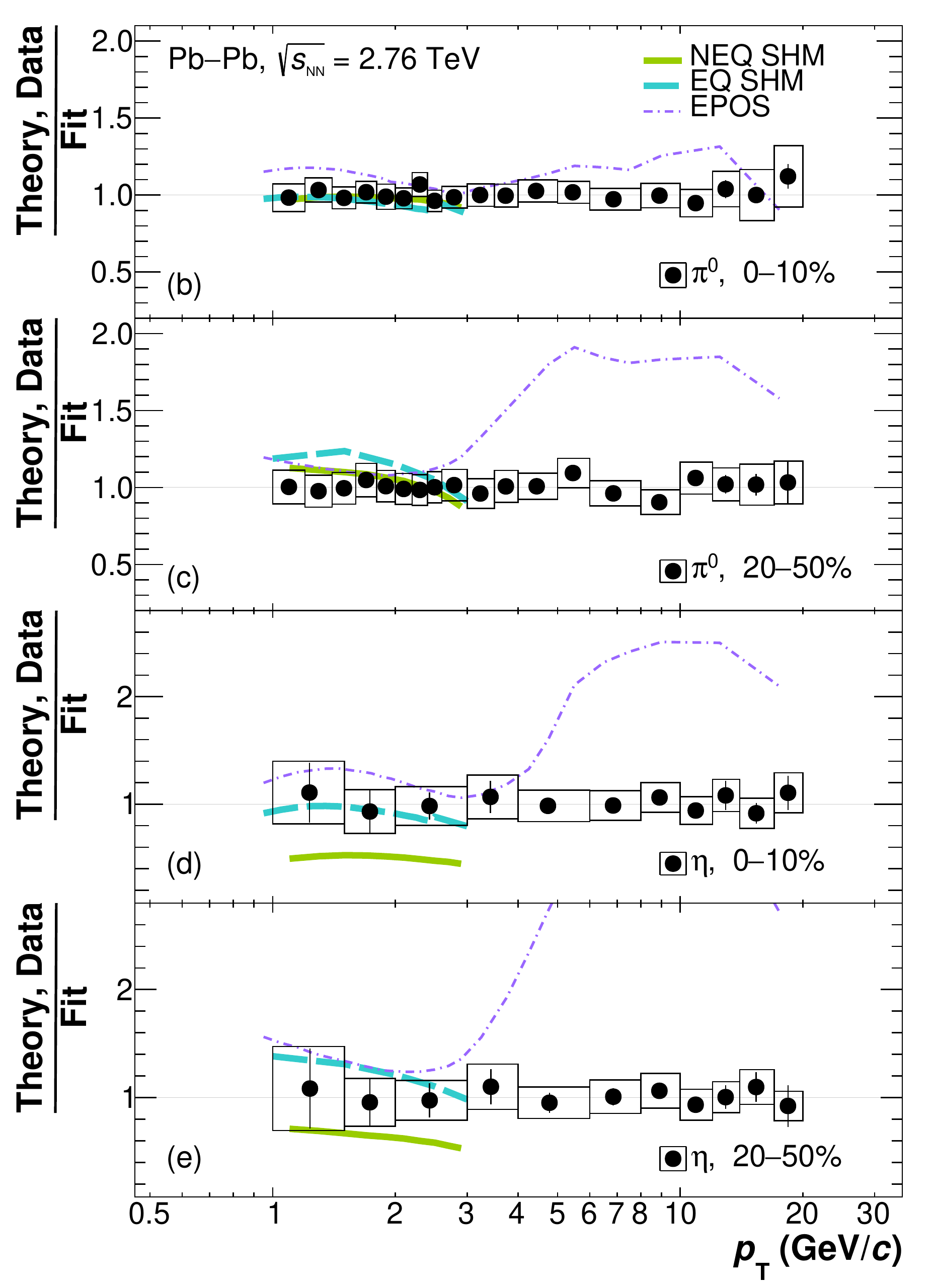}
\caption{(Color online) (a) Comparison of the \pai (open symbols) and \e (closed symbols) 
meson invariant \pT-differential yields to EQ and NEQ SHM~\cite{Begun:2013nga} and to 
EPOS~\cite{PhysRevC.85.064907} for the two centrality classes measured. (b$-$e) Ratio of data 
and theory calculations to the fit of the \pai and \e invariant yields (see Fig.~\ref{fig:InvYieldsDataOnly} 
and the left plot of this figure) in the two centrality classes measured. Black points 
show the data to fit ratio (vertical lines for statistical errors and boxes for systematic 
errors). The bold lines correspond to the ratio of the EQ and NEQ SHM predictions~\cite{Begun:2013nga} 
to the data fit, while the thin dashed lines correspond to the same ratio for the EPOS 
predictions~\cite{PhysRevC.85.064907}.} 
\label{fig:begun}
\end{figure}

\begin{figure}[!h]
\centering
\includegraphics[width=0.5\textwidth]{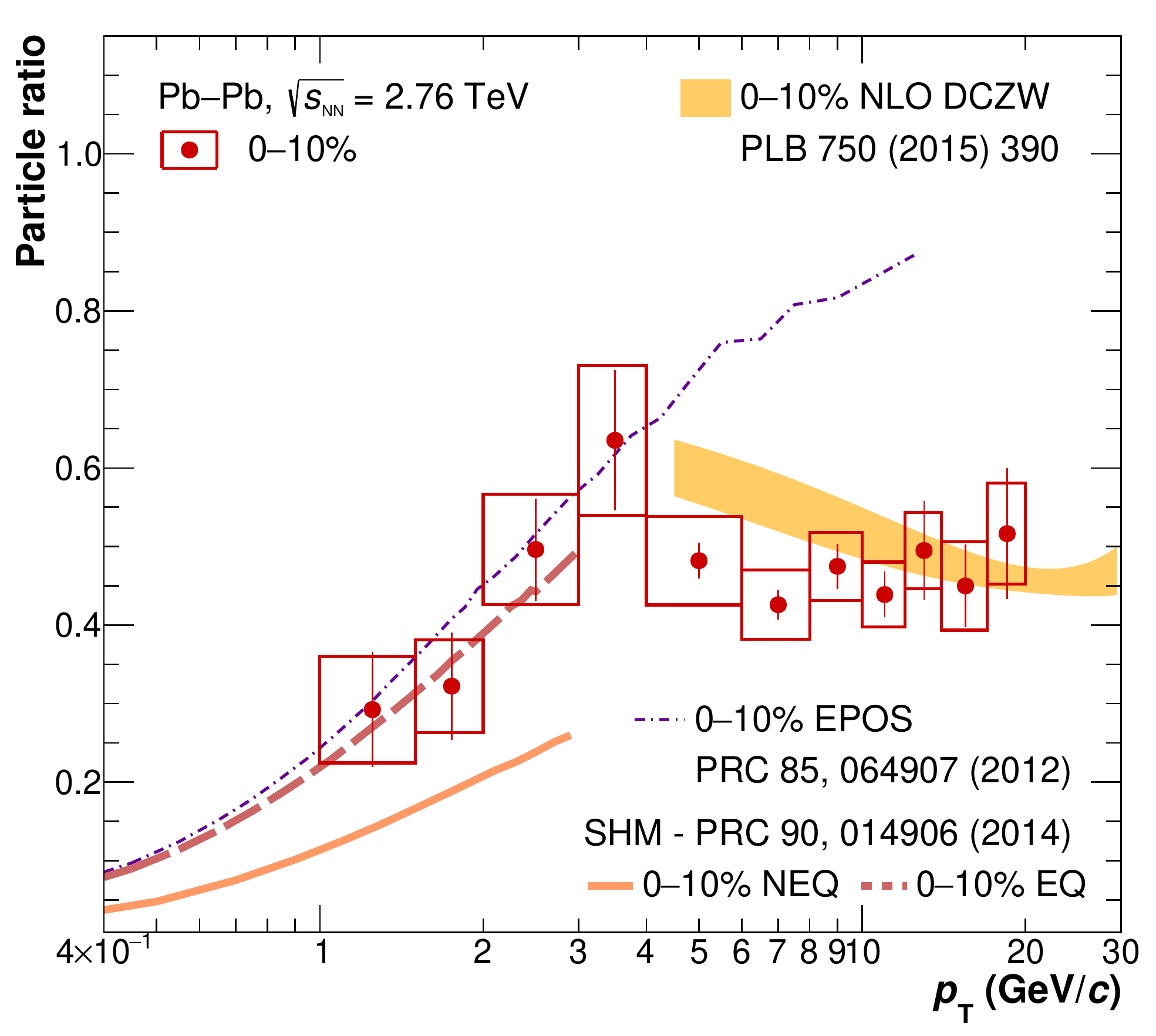}
\caption{(Color online) Comparison of the \etapai ratio in the centrality class 0--10\% 
(circles) to the NLO pQCD prediction by DCZW (solid band)~\cite{Dai:2015dxa} and to the 
EQ and NEQ SHM~\cite{Begun:2013nga} and EPOS~\cite{PhysRevC.85.064907} predictions for 
the input yields (bold and thin dashed lines respectively).} 
\label{fig:etatopi0Theory}
\end{figure}

\begin{figure}[h]
\centering
\includegraphics[width=0.85\textwidth]{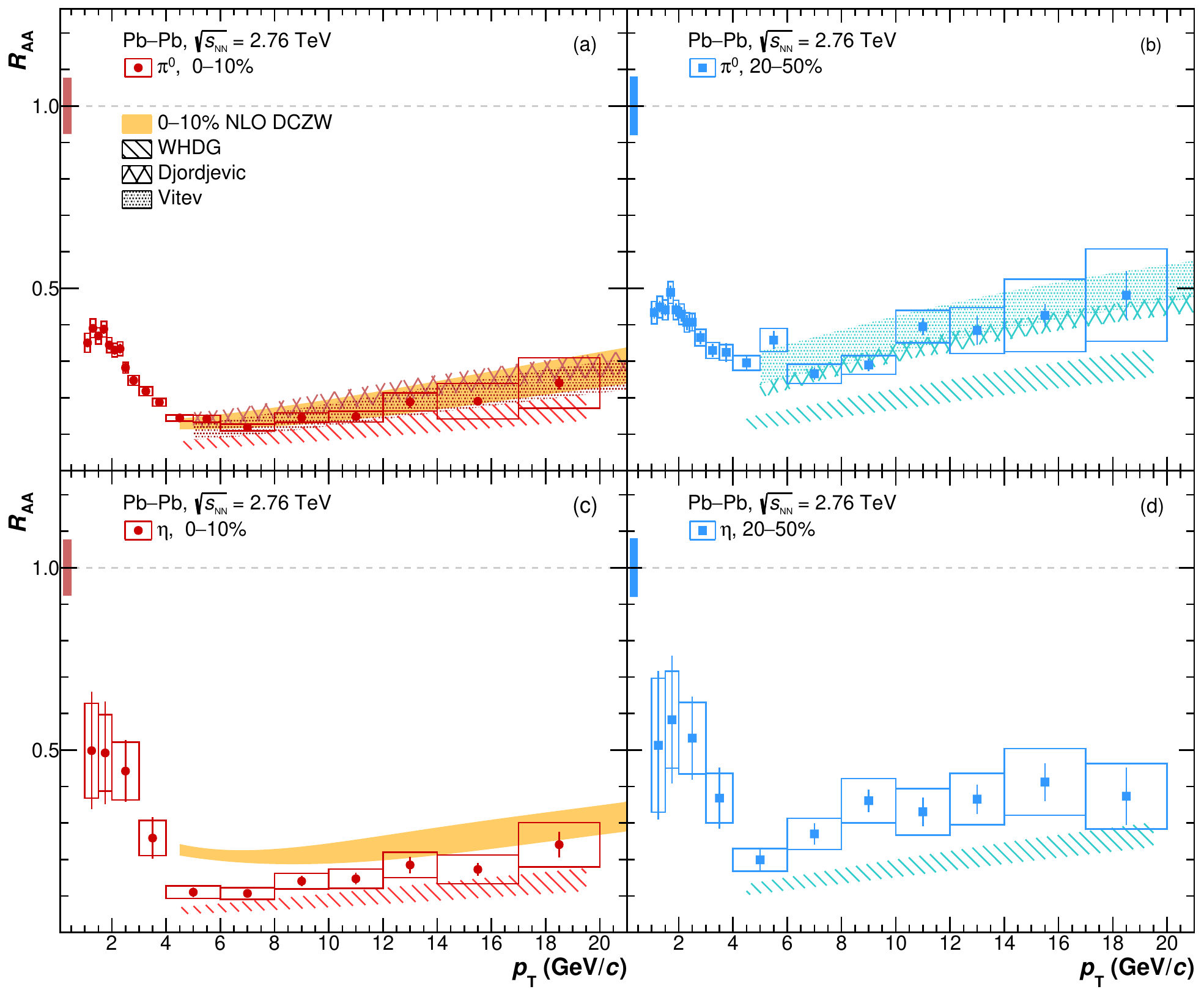}
\caption{(Color online) \RAA of (a, b) \pai and (c, d) \e meson compared to NLO pQCD 
predictions by DCZW (solid bands)~\cite{Dai:2015dxa}, by WHDG (dashed bands)~\cite{Horowitz:2007nq}, 
Djordevic~\emph{et al.}~\cite{Djordjevic2014286} (crossed bands, \pai only) and 
Vitev~\emph{et al.}~\cite{Vitev:2005he,Kang:2014xsa,Chien:2015vja,Bauer:2010cc} (empty bands, \pai only) 
in the two centrality classes measured.}
\label{fig:RAATh}
\end{figure}

The \pai and \e invariant \pT-differential yields are compared to predictions using a 
statistical hadronization model (SHM)~\cite{Begun:2013nga,Broniowski:2001we} and the 
EPOS2~\cite{PhysRevC.85.064907} event generator. Results from two versions of the SHM are 
presented here, an equilibrium (EQ) and non-equilibrium (NEQ) prediction. In the NEQ SHM, 
the mean particle multiplicities are described with the use of four thermodynamic parameters: 
temperature $T$, volume $V$, and two parameters to account for the non-equilibrium conditions 
-- $\gamma_s$ and $\gamma_q$. The EQ SHM can be treated as a particular case of the NEQ when 
$\gamma_s $~=~$\gamma_q$ = 1. The parameters of the model are determined by fits to the 
measured charged pion and kaon spectra~\cite{Begun:2013nga}. While only these two particles 
are considered in the fits, the resulting parameters are used to make predictions for other 
particles~\cite{Begun:2017hpi}, \emph{e.g.} the \e meson, the $\rho$ meson and the proton. 
The EPOS generator addresses both low- and high-\pT phenomena, where the particle spectra 
include effects (low-\pT) associated to hydrodynamic flow as discussed in \cite{Abelev:2013vea}. 
At higher \pT, the focus is shifted towards energy loss of high-\pT strings where strings 
are the byproduct of hard scatterings. \\
Fig.~\ref{fig:begun} (a) shows the comparison to models for the 0--10\% and 20--50\% centrality 
classes while Fig.~\ref{fig:begun} (b$-$e) shows the ratio of data and theory calculations 
to the fit of the \pai  and \e invariant yields. 
\\
The EQ and NEQ SHM predictions (bold lines in Fig.~\ref{fig:begun}, (b) and (c)) describe 
the shape of the \pai measurement within the uncertainties for both centralities. For the 
\e meson, in (d) and (e), the EQ model also describes the data within uncertainties. Conversely, the NEQ model 
predicts about half as many $\eta$ mesons than actually measured in central collisions. 
The difference observed between the NEQ SHM and the data may point towards a different 
flow profile of the two mesons with a larger flow for the \e than the \pai~\cite{Retiere:2003kf}.
Significant differences between the EQ and NEQ predictions are also observed for the 
$\rho^{0}$, $\Sigma$(1385), $\Lambda$(1520) and $\Xi$(1530)~\cite{Begun:2014rsa,Begun:2017hpi}. \\
The \pai and \e mesons are only partially described by EPOS (dashed lines in Fig.~\ref{fig:begun}, 
(b$-$e)). While the comparison is reasonably close to the data points for the \pai measurement 
in 0--10\% (b), the model only describes the low \pT part of the (c) semicentral \pai and 
(d, e) \e measurements. No theoretical uncertainties for the EPOS calculations are available 
at the time of writing. \\
The \etapai ratio for the centrality class 0--10\% is compared to the NLO pQCD calculation 
by DCZW (Dai, Chen, Zhang, and Wang)~\cite{Dai:2015dxa}, to the ratio from the EQ and NEQ SHM~\cite{Begun:2013nga} 
predictions and the EPOS~\cite{PhysRevC.85.064907} generator in Fig.~\ref{fig:etatopi0Theory}. 
The DCZW model is based on a higher twist approach to jet quenching~\cite{Guo:2000nz} where 
parton fragmentation functions are modified as a consequence of the parton energy loss. 
A generalized QCD factorization of twist-4 processes is used to calculate the scattering. 
The effective parton fragmentation functions AKK (Albino, Kniehl, Kramer) ~\cite{Albino:2005me} and AESS (Aidala, Ellinghaus, Seele, Stratmann)~\cite{Aidala:2010bn}) 
are then incorporated into a NLO pQCD framework to describe the particle production 
suppression. Data and the DCZW prediction are in agreement within uncertainties. The EQ 
SHM prediction describes the \etapai ratio, while in comparison to the NEQ SHM prediction 
the ratio is underestimated as shown in Fig.~\ref{fig:begun} (d, e). The EPOS curves 
describe the ratio up to 4~\GeVc, as expected since the disagreement with the \e meson 
measurement is larger at higher \pT. \\
The measurements of \RAA for both mesons are compared to four NLO pQCD based models in 
Fig.~\ref{fig:RAATh}: DCZW~\cite{Dai:2015dxa} (described above), WHDG(Wicks, Horowitz, Djordjevic and Gyulassy)~\cite{Wicks:2005gt,Horowitz:2011gd,Horowitz:2007nq}, 
Djordevic~\emph{et al.}~\cite{Djordjevic2014286} (\pai only) and Vitev~\emph{et al.}~\cite{Vitev:2005he,Kang:2014xsa,Chien:2015vja,Bauer:2010cc} 
(\pai only). In the first three models, it is assumed that a fast moving parton passing 
through hot partonic matter will lose its energy via induced radiation due to multiple 
parton scattering. The WHDG calculation models collisional and radiative energy loss 
processes in a Bjorken-expanding medium. It assumes that the color charge density of 
the medium is proportional to the number of participating nucleons obtained from a 
Glauber model. Hard parton-parton scatterings are then proportional to the number of 
binary nucleon-nucleon collisions. The Djordevic~\emph{et al.} model also includes effects 
due to the finite size of the QCD medium, the finite magnetic mass and the running of 
the coupling~\cite{PhysRevC.74.064907,PhysRevLett.101.022302,Djordjevic2012229,Djordjevic2014286}. 
The model of Vitev~\emph{et al.} is an application of the soft-collinear effective theory 
with Glauber gluons (SCET$_\mathrm{G}$) to study inclusive hadron suppression in 
nucleus-nucleus collisions. In this model, medium-evolved fragmentation functions are 
combined with all initial-state cold nuclear matter (CNM) effects (dynamical nuclear 
shadowing, Cronin effect and initial-state parton energy loss). The authors demonstrate 
that traditional parton energy loss calculations can be regarded as a special soft-gluon 
emission limit of the general QCD evolution framework. \\ 
In the most central event class, the \pai meson \RAA is described for \pT~$>$~4~\GeVc 
by the DCZW, Djordevic~\emph{et al.} and Vitev~\emph{et al.} models and for \pT~$>$~6~\GeVc 
by WHDG (Fig.~\ref{fig:RAATh}, (a)). For the DCZW predictions, the \e meson is described 
within uncertainties from \pT~$>$~8~\GeVc; below this momentum, the DCZW model overstimates 
the \RAA result (Fig.~\ref{fig:RAATh}, (c)). The latter may indicate that the relative 
quark and gluon contributions to the \e meson production is overestimated at intermediate 
\pT (4~$<$~\pT~$<$~8~\GeVc). On the other hand, the WHDG model predicts larger suppression 
than observed in the data for the \e meson in the centrality class 0--10\% and for both 
mesons in the centrality class 20--50\%. The Djordevic~\emph{et al.} and Vitev~\emph{et al.} 
models describe the \pai meson suppression in both centrality classes within uncertainties.

\section{Summary}

We have presented measurements of the \pai and \e meson production at mid-rapidity in \Pb 
collisions at \sNN measured with the ALICE detector. Independent and complementary 
techniques are used: photon detection with electromagnetic calorimetry and photon 
reconstruction through conversions using the tracking system. The combination of these 
methods allowed measurements in a large transverse momentum range, from 1 to 20~\GeVc.\\
The results represent the first measurement of \e meson production in heavy-ion collisions 
at the LHC. The \pai measurements are performed using data that corresponds to a factor 10 
increase in integrated luminosity with respect to the previous ALICE publication~\cite{Abelev:2014ypa}. 
The higher statistics allowed for an improved measurement that probes the \pT region up 
to 20~\GeVc. 

The \etapai ratio is compared to NLO pQCD calculations, corresponding ALICE measurements 
in pp collisions and to the $K^{\pm}/\pi^{\pm}$ ratio measured in \Pb collisions at the same 
energy. For \pT~$>$~4~\GeVc, these results indicate that the ratio in \Pb is similar to the 
vacuum expectation, assuming this to be the pp measurement. The ratio is also consistent 
with predictions from pQCD-based calculations within experimental uncertainties. No effects 
beyond one $\sigma$ related to the strange quark content, mass hierarchy between particles 
or contributions from higher mass resonance decays that may lead to discernible differences 
between \etapai and $K^{\pm}/\pi^{\pm}$ were observed.

The invariant yields of both mesons as well as the \etapai ratio are compared to predictions 
including a hydrodynamic approach focusing on low-\pT phenomena. These comparisons show 
different levels of agreement for \e and \pai. EPOS slightly overestimates the production rates 
of the two mesons at low-\pT, but shows a much larger deviation above 3--4~\GeVc. Both the 
EQ and NEQ SHM predictions describe the measured \pai production rates. The data favors the 
EQ model description which agrees with the \e measurement. The NEQ model is disfavored by 
the data as it underestimates the results by a factor of two.

The RAA results show an increasing trend at high \pT which may be explained by a larger quark 
to gluon contribution in the production of neutral mesons. The \RAA \pai measurements, when 
compared to world data, confirm the center-of-mass energy dependence of the observed 
suppression when going from low (SPS) to higher (RHIC, ALICE) collision energies. Results 
of \RAA for \e mesons are currently available only at two center-of-mass energies from the LHC 
and RHIC with sizable uncertainties. Due to the lack of precise world data, it is difficult to 
conclude on an energy dependence of the \e suppression.

The \RAA results are additionally compared to NLO pQCD calculations. The WHDG model describes the 
suppression observed for the \pai meson in the 0--10$\%$ centrality class within theoretical and 
experimental uncertainties. For the \e measurement, the model predicts a larger suppression than 
observed. In the 20--50\% centrality class the predictions are in disagreement by several sigma 
with the ALICE data for both mesons. The DCZW model describes within uncertainties the \pai 
measurement and the \e meson above 8~\GeVc. Below this \pT, the model predicts less suppression 
than observed. The Djordevic~\emph{et al.} and Vitev~\emph{et al.} calculations describe well 
the \pai production rates in both centrality classes. The disagreement observed between the \e 
measurements and the models may point to a overestimation (DCZW) or underestimation (WHDG) of 
the gluon to quark contributions to the \e meson production in heavy-ion collisions at LHC energies.  

The presented results, when compared to models, highlight the lack of a full theoretical 
description of neutral meson production. The measurements presented in this paper will be 
essential to further constrain theoretical models and improve our understanding of the 
experimental results.

\newenvironment{acknowledgement}{\relax}{\relax}
\begin{acknowledgement}
\section*{Acknowledgements}
We would like to thank Viktor Begun, Wei Dai, Magdalena Djordevic, William Horowitz, Marco 
Stratmann, Ivan Vitev and Klaus Werner for useful discussions and clarifications, and for 
providing the predictions shown in this paper.


The ALICE Collaboration would like to thank all its engineers and technicians for their invaluable contributions to the construction of the experiment and the CERN accelerator teams for the outstanding performance of the LHC complex.
The ALICE Collaboration gratefully acknowledges the resources and support provided by all Grid centres and the Worldwide LHC Computing Grid (WLCG) collaboration.
The ALICE Collaboration acknowledges the following funding agencies for their support in building and running the ALICE detector:
A. I. Alikhanyan National Science Laboratory (Yerevan Physics Institute) Foundation (ANSL), State Committee of Science and World Federation of Scientists (WFS), Armenia;
Austrian Academy of Sciences and Nationalstiftung f\"{u}r Forschung, Technologie und Entwicklung, Austria;
Ministry of Communications and High Technologies, National Nuclear Research Center, Azerbaijan;
Conselho Nacional de Desenvolvimento Cient\'{\i}fico e Tecnol\'{o}gico (CNPq), Universidade Federal do Rio Grande do Sul (UFRGS), Financiadora de Estudos e Projetos (Finep) and Funda\c{c}\~{a}o de Amparo \`{a} Pesquisa do Estado de S\~{a}o Paulo (FAPESP), Brazil;
Ministry of Science \& Technology of China (MSTC), National Natural Science Foundation of China (NSFC) and Ministry of Education of China (MOEC) , China;
Ministry of Science, Education and Sport and Croatian Science Foundation, Croatia;
Ministry of Education, Youth and Sports of the Czech Republic, Czech Republic;
The Danish Council for Independent Research | Natural Sciences, the Carlsberg Foundation and Danish National Research Foundation (DNRF), Denmark;
Helsinki Institute of Physics (HIP), Finland;
Commissariat \`{a} l'Energie Atomique (CEA) and Institut National de Physique Nucl\'{e}aire et de Physique des Particules (IN2P3) and Centre National de la Recherche Scientifique (CNRS), France;
Bundesministerium f\"{u}r Bildung, Wissenschaft, Forschung und Technologie (BMBF) and GSI Helmholtzzentrum f\"{u}r Schwerionenforschung GmbH, Germany;
General Secretariat for Research and Technology, Ministry of Education, Research and Religions, Greece;
National Research, Development and Innovation Office, Hungary;
Department of Atomic Energy Government of India (DAE), Department of Science and Technology, Government of India (DST), University Grants Commission, Government of India (UGC) and Council of Scientific and Industrial Research (CSIR), India;
Indonesian Institute of Science, Indonesia;
Centro Fermi - Museo Storico della Fisica e Centro Studi e Ricerche Enrico Fermi and Istituto Nazionale di Fisica Nucleare (INFN), Italy;
Institute for Innovative Science and Technology , Nagasaki Institute of Applied Science (IIST), Japan Society for the Promotion of Science (JSPS) KAKENHI and Japanese Ministry of Education, Culture, Sports, Science and Technology (MEXT), Japan;
Consejo Nacional de Ciencia (CONACYT) y Tecnolog\'{i}a, through Fondo de Cooperaci\'{o}n Internacional en Ciencia y Tecnolog\'{i}a (FONCICYT) and Direcci\'{o}n General de Asuntos del Personal Academico (DGAPA), Mexico;
Nederlandse Organisatie voor Wetenschappelijk Onderzoek (NWO), Netherlands;
The Research Council of Norway, Norway;
Commission on Science and Technology for Sustainable Development in the South (COMSATS), Pakistan;
Pontificia Universidad Cat\'{o}lica del Per\'{u}, Peru;
Ministry of Science and Higher Education and National Science Centre, Poland;
Korea Institute of Science and Technology Information and National Research Foundation of Korea (NRF), Republic of Korea;
Ministry of Education and Scientific Research, Institute of Atomic Physics and Romanian National Agency for Science, Technology and Innovation, Romania;
Joint Institute for Nuclear Research (JINR), Ministry of Education and Science of the Russian Federation and National Research Centre Kurchatov Institute, Russia;
Ministry of Education, Science, Research and Sport of the Slovak Republic, Slovakia;
National Research Foundation of South Africa, South Africa;
%
%
Swedish Research Council (VR) and Knut \& Alice Wallenberg Foundation (KAW), Sweden;
European Organization for Nuclear Research, Switzerland;
National Science and Technology Development Agency (NSDTA), Suranaree University of Technology (SUT) and Office of the Higher Education Commission under NRU project of Thailand, Thailand;
Turkish Atomic Energy Agency (TAEK), Turkey;
National Academy of  Sciences of Ukraine, Ukraine;
Science and Technology Facilities Council (STFC), United Kingdom;
National Science Foundation of the United States of America (NSF) and United States Department of Energy, Office of Nuclear Physics (DOE NP), United States of America.
 
\end{acknowledgement}

\bibliographystyle{utphys}   

\bibliography{bibliography.bib}{}
\newpage
\appendix
\section{The ALICE Collaboration}
\label{app:collab}

\begingroup
\small
\begin{flushleft}
S.~Acharya\Irefn{org138}\And 
F.T.-.~Acosta\Irefn{org22}\And 
D.~Adamov\'{a}\Irefn{org94}\And 
J.~Adolfsson\Irefn{org81}\And 
M.M.~Aggarwal\Irefn{org98}\And 
G.~Aglieri Rinella\Irefn{org36}\And 
M.~Agnello\Irefn{org33}\And 
N.~Agrawal\Irefn{org48}\And 
Z.~Ahammed\Irefn{org138}\And 
S.U.~Ahn\Irefn{org77}\And 
S.~Aiola\Irefn{org143}
% ...
\And
\renewcommand\labelenumi{\textsuperscript{\theenumi}~}

\section*{Affiliation notes}
\renewcommand\theenumi{\roman{enumi}}
\begin{Authlist}
\item \Adef{org*}Deceased
\item \Adef{orgI}Dipartimento DET del Politecnico di Torino, Turin, Italy
\item \Adef{orgII}M.V. Lomonosov Moscow State University, D.V. Skobeltsyn Institute of Nuclear, Physics, Moscow, Russia
\item \Adef{orgIII}Department of Applied Physics, Aligarh Muslim University, Aligarh, India
\item \Adef{orgIV}Institute of Theoretical Physics, University of Wroclaw, Poland
\end{Authlist}

\section*{Collaboration Institutes}
\renewcommand\theenumi{\arabic{enumi}~}
\begin{Authlist}
\item \Idef{org1}A.I. Alikhanyan National Science Laboratory (Yerevan Physics Institute) Foundation, Yerevan, Armenia
\item \Idef{org2}Benem\'{e}rita Universidad Aut\'{o}noma de Puebla, Puebla, Mexico
\item \Idef{org3}Bogolyubov Institute for Theoretical Physics, National Academy of Sciences of Ukraine, Kiev, Ukraine
\item \Idef{org4}Bose Institute, Department of Physics  and Centre for Astroparticle Physics and Space Science (CAPSS), Kolkata, India
\item \Idef{org5}Budker Institute for Nuclear Physics, Novosibirsk, Russia
\item \Idef{org6}California Polytechnic State University, San Luis Obispo, California, United States
\item \Idef{org7}Central China Normal University, Wuhan, China
\item \Idef{org8}Centre de Calcul de l'IN2P3, Villeurbanne, Lyon, France
\item \Idef{org9}Centro de Aplicaciones Tecnol\'{o}gicas y Desarrollo Nuclear (CEADEN), Havana, Cuba
\item \Idef{org10}Centro de Investigaci\'{o}n y de Estudios Avanzados (CINVESTAV), Mexico City and M\'{e}rida, Mexico
\item \Idef{org11}Centro Fermi - Museo Storico della Fisica e Centro Studi e Ricerche ``Enrico Fermi', Rome, Italy
\item \Idef{org12}Chicago State University, Chicago, Illinois, United States
\item \Idef{org13}China Institute of Atomic Energy, Beijing, China
\item \Idef{org14}Chonbuk National University, Jeonju, Republic of Korea
\item \Idef{org15}Comenius University Bratislava, Faculty of Mathematics, Physics and Informatics, Bratislava, Slovakia
\item \Idef{org16}COMSATS Institute of Information Technology (CIIT), Islamabad, Pakistan
\item \Idef{org17}Creighton University, Omaha, Nebraska, United States
\item \Idef{org18}Department of Physics, Aligarh Muslim University, Aligarh, India
\item \Idef{org19}Department of Physics, Ohio State University, Columbus, Ohio, United States
\item \Idef{org20}Department of Physics, Pusan National University, Pusan, Republic of Korea
\item \Idef{org21}Department of Physics, Sejong University, Seoul, Republic of Korea
\item \Idef{org22}Department of Physics, University of California, Berkeley, California, United States
\item \Idef{org23}Department of Physics, University of Oslo, Oslo, Norway
\item \Idef{org24}Department of Physics and Technology, University of Bergen, Bergen, Norway
\item \Idef{org25}Dipartimento di Fisica dell'Universit\`{a} 'La Sapienza' and Sezione INFN, Rome, Italy
\item \Idef{org26}Dipartimento di Fisica dell'Universit\`{a} and Sezione INFN, Cagliari, Italy
\item \Idef{org27}Dipartimento di Fisica dell'Universit\`{a} and Sezione INFN, Trieste, Italy
\item \Idef{org28}Dipartimento di Fisica dell'Universit\`{a} and Sezione INFN, Turin, Italy
\item \Idef{org29}Dipartimento di Fisica e Astronomia dell'Universit\`{a} and Sezione INFN, Bologna, Italy
\item \Idef{org30}Dipartimento di Fisica e Astronomia dell'Universit\`{a} and Sezione INFN, Catania, Italy
\item \Idef{org31}Dipartimento di Fisica e Astronomia dell'Universit\`{a} and Sezione INFN, Padova, Italy
\item \Idef{org32}Dipartimento di Fisica `E.R.~Caianiello' dell'Universit\`{a} and Gruppo Collegato INFN, Salerno, Italy
\item \Idef{org33}Dipartimento DISAT del Politecnico and Sezione INFN, Turin, Italy
\item \Idef{org34}Dipartimento di Scienze e Innovazione Tecnologica dell'Universit\`{a} del Piemonte Orientale and INFN Sezione di Torino, Alessandria, Italy
\item \Idef{org35}Dipartimento Interateneo di Fisica `M.~Merlin' and Sezione INFN, Bari, Italy
\item \Idef{org36}European Organization for Nuclear Research (CERN), Geneva, Switzerland
\item \Idef{org37}Faculty of Engineering and Business Administration, Western Norway University of Applied Sciences, Bergen, Norway
\item \Idef{org38}Faculty of Nuclear Sciences and Physical Engineering, Czech Technical University in Prague, Prague, Czech Republic
\item \Idef{org39}Faculty of Science, P.J.~\v{S}af\'{a}rik University, Ko\v{s}ice, Slovakia
\item \Idef{org40}Frankfurt Institute for Advanced Studies, Johann Wolfgang Goethe-Universit\"{a}t Frankfurt, Frankfurt, Germany
\item \Idef{org41}Gangneung-Wonju National University, Gangneung, Republic of Korea
\item \Idef{org42}Gauhati University, Department of Physics, Guwahati, India
\item \Idef{org43}Helmholtz-Institut f\"{u}r Strahlen- und Kernphysik, Rheinische Friedrich-Wilhelms-Universit\"{a}t Bonn, Bonn, Germany
\item \Idef{org44}Helsinki Institute of Physics (HIP), Helsinki, Finland
\item \Idef{org45}Hiroshima University, Hiroshima, Japan
\item \Idef{org46}Hochschule Worms, Zentrum  f\"{u}r Technologietransfer und Telekommunikation (ZTT), Worms, Germany
\item \Idef{org47}Horia Hulubei National Institute of Physics and Nuclear Engineering, Bucharest, Romania
\item \Idef{org48}Indian Institute of Technology Bombay (IIT), Mumbai, India
\item \Idef{org49}Indian Institute of Technology Indore, Indore, India
\item \Idef{org50}Indonesian Institute of Sciences, Jakarta, Indonesia
\item \Idef{org51}INFN, Laboratori Nazionali di Frascati, Frascati, Italy
\item \Idef{org52}INFN, Sezione di Bari, Bari, Italy
\item \Idef{org53}INFN, Sezione di Bologna, Bologna, Italy
\item \Idef{org54}INFN, Sezione di Cagliari, Cagliari, Italy
\item \Idef{org55}INFN, Sezione di Catania, Catania, Italy
\item \Idef{org56}INFN, Sezione di Padova, Padova, Italy
\item \Idef{org57}INFN, Sezione di Roma, Rome, Italy
\item \Idef{org58}INFN, Sezione di Torino, Turin, Italy
\item \Idef{org59}INFN, Sezione di Trieste, Trieste, Italy
\item \Idef{org60}Inha University, Incheon, Republic of Korea
\item \Idef{org61}Institut de Physique Nucl\'{e}aire d'Orsay (IPNO), Institut National de Physique Nucl\'{e}aire et de Physique des Particules (IN2P3/CNRS), Universit\'{e} de Paris-Sud, Universit\'{e} Paris-Saclay, Orsay, France
\item \Idef{org62}Institute for Nuclear Research, Academy of Sciences, Moscow, Russia
\item \Idef{org63}Institute for Subatomic Physics of Utrecht University, Utrecht, Netherlands
\item \Idef{org64}Institute for Theoretical and Experimental Physics, Moscow, Russia
\item \Idef{org65}Institute of Experimental Physics, Slovak Academy of Sciences, Ko\v{s}ice, Slovakia
\item \Idef{org66}Institute of Physics, Bhubaneswar, India
\item \Idef{org67}Institute of Physics of the Czech Academy of Sciences, Prague, Czech Republic
\item \Idef{org68}Institute of Space Science (ISS), Bucharest, Romania
\item \Idef{org69}Institut f\"{u}r Kernphysik, Johann Wolfgang Goethe-Universit\"{a}t Frankfurt, Frankfurt, Germany
\item \Idef{org70}Instituto de Ciencias Nucleares, Universidad Nacional Aut\'{o}noma de M\'{e}xico, Mexico City, Mexico
\item \Idef{org71}Instituto de F\'{i}sica, Universidade Federal do Rio Grande do Sul (UFRGS), Porto Alegre, Brazil
\item \Idef{org72}Instituto de F\'{\i}sica, Universidad Nacional Aut\'{o}noma de M\'{e}xico, Mexico City, Mexico
\item \Idef{org73}iThemba LABS, National Research Foundation, Somerset West, South Africa
\item \Idef{org74}Johann-Wolfgang-Goethe Universit\"{a}t Frankfurt Institut f\"{u}r Informatik, Fachbereich Informatik und Mathematik, Frankfurt, Germany
\item \Idef{org75}Joint Institute for Nuclear Research (JINR), Dubna, Russia
\item \Idef{org76}Konkuk University, Seoul, Republic of Korea
\item \Idef{org77}Korea Institute of Science and Technology Information, Daejeon, Republic of Korea
\item \Idef{org78}KTO Karatay University, Konya, Turkey
\item \Idef{org79}Laboratoire de Physique Subatomique et de Cosmologie, Universit\'{e} Grenoble-Alpes, CNRS-IN2P3, Grenoble, France
\item \Idef{org80}Lawrence Berkeley National Laboratory, Berkeley, California, United States
\item \Idef{org81}Lund University Department of Physics, Division of Particle Physics, Lund, Sweden
\item \Idef{org82}Nagasaki Institute of Applied Science, Nagasaki, Japan
\item \Idef{org83}Nara Women{'}s University (NWU), Nara, Japan
\item \Idef{org84}National and Kapodistrian University of Athens, School of Science, Department of Physics , Athens, Greece
\item \Idef{org85}National Centre for Nuclear Research, Warsaw, Poland
\item \Idef{org86}National Institute of Science Education and Research, HBNI, Jatni, India
\item \Idef{org87}National Nuclear Research Center, Baku, Azerbaijan
\item \Idef{org88}National Research Centre Kurchatov Institute, Moscow, Russia
\item \Idef{org89}Niels Bohr Institute, University of Copenhagen, Copenhagen, Denmark
\item \Idef{org90}Nikhef, National institute for subatomic physics, Amsterdam, Netherlands
\item \Idef{org91}NRC ¿Kurchatov Institute¿ ¿ IHEP , Protvino, Russia
\item \Idef{org92}NRNU Moscow Engineering Physics Institute, Moscow, Russia
\item \Idef{org93}Nuclear Physics Group, STFC Daresbury Laboratory, Daresbury, United Kingdom
\item \Idef{org94}Nuclear Physics Institute of the Czech Academy of Sciences, \v{R}e\v{z} u Prahy, Czech Republic
\item \Idef{org95}Oak Ridge National Laboratory, Oak Ridge, Tennessee, United States
\item \Idef{org96}Petersburg Nuclear Physics Institute, Gatchina, Russia
\item \Idef{org97}Physics department, Faculty of science, University of Zagreb, Zagreb, Croatia
\item \Idef{org98}Physics Department, Panjab University, Chandigarh, India
\item \Idef{org99}Physics Department, University of Jammu, Jammu, India
\item \Idef{org100}Physics Department, University of Rajasthan, Jaipur, India
\item \Idef{org101}Physikalisches Institut, Eberhard-Karls-Universit\"{a}t T\"{u}bingen, T\"{u}bingen, Germany
\item \Idef{org102}Physikalisches Institut, Ruprecht-Karls-Universit\"{a}t Heidelberg, Heidelberg, Germany
\item \Idef{org103}Physik Department, Technische Universit\"{a}t M\"{u}nchen, Munich, Germany
\item \Idef{org104}Research Division and ExtreMe Matter Institute EMMI, GSI Helmholtzzentrum f\"ur Schwerionenforschung GmbH, Darmstadt, Germany
\item \Idef{org105}Rudjer Bo\v{s}kovi\'{c} Institute, Zagreb, Croatia
\item \Idef{org106}Russian Federal Nuclear Center (VNIIEF), Sarov, Russia
\item \Idef{org107}Saha Institute of Nuclear Physics, Kolkata, India
\item \Idef{org108}School of Physics and Astronomy, University of Birmingham, Birmingham, United Kingdom
\item \Idef{org109}Secci\'{o}n F\'{\i}sica, Departamento de Ciencias, Pontificia Universidad Cat\'{o}lica del Per\'{u}, Lima, Peru
\item \Idef{org110}Stefan Meyer Institut f\"{u}r Subatomare Physik (SMI), Vienna, Austria
\item \Idef{org111}SUBATECH, IMT Atlantique, Universit\'{e} de Nantes, CNRS-IN2P3, Nantes, France
\item \Idef{org112}Suranaree University of Technology, Nakhon Ratchasima, Thailand
\item \Idef{org113}Technical University of Ko\v{s}ice, Ko\v{s}ice, Slovakia
\item \Idef{org114}Technische Universit\"{a}t M\"{u}nchen, Excellence Cluster 'Universe', Munich, Germany
\item \Idef{org115}The Henryk Niewodniczanski Institute of Nuclear Physics, Polish Academy of Sciences, Cracow, Poland
\item \Idef{org116}The University of Texas at Austin, Austin, Texas, United States
\item \Idef{org117}Universidad Aut\'{o}noma de Sinaloa, Culiac\'{a}n, Mexico
\item \Idef{org118}Universidade de S\~{a}o Paulo (USP), S\~{a}o Paulo, Brazil
\item \Idef{org119}Universidade Estadual de Campinas (UNICAMP), Campinas, Brazil
\item \Idef{org120}Universidade Federal do ABC, Santo Andre, Brazil
\item \Idef{org121}University College of Southeast Norway, Tonsberg, Norway
\item \Idef{org122}University of Cape Town, Cape Town, South Africa
\item \Idef{org123}University of Houston, Houston, Texas, United States
\item \Idef{org124}University of Jyv\"{a}skyl\"{a}, Jyv\"{a}skyl\"{a}, Finland
\item \Idef{org125}University of Liverpool, Liverpool, United Kingdom
\item \Idef{org126}University of Split, Faculty of Electrical Engineering, Mechanical Engineering and Naval Architecture, Split, Croatia
\item \Idef{org127}University of Tennessee, Knoxville, Tennessee, United States
\item \Idef{org128}University of the Witwatersrand, Johannesburg, South Africa
\item \Idef{org129}University of Tokyo, Tokyo, Japan
\item \Idef{org130}University of Tsukuba, Tsukuba, Japan
\item \Idef{org131}Universit\'{e} Clermont Auvergne, CNRS/IN2P3, LPC, Clermont-Ferrand, France
\item \Idef{org132}Universit\'{e} de Lyon, Universit\'{e} Lyon 1, CNRS/IN2P3, IPN-Lyon, Villeurbanne, Lyon, France
\item \Idef{org133}Universit\'{e} de Strasbourg, CNRS, IPHC UMR 7178, F-67000 Strasbourg, France, Strasbourg, France
\item \Idef{org134} Universit\'{e} Paris-Saclay Centre d¿\'Etudes de Saclay (CEA), IRFU, Department de Physique Nucl\'{e}aire (DPhN), Saclay, France
\item \Idef{org135}Universit\`{a} degli Studi di Pavia, Pavia, Italy
\item \Idef{org136}Universit\`{a} di Brescia, Brescia, Italy
\item \Idef{org137}V.~Fock Institute for Physics, St. Petersburg State University, St. Petersburg, Russia
\item \Idef{org138}Variable Energy Cyclotron Centre, Kolkata, India
\item \Idef{org139}Warsaw University of Technology, Warsaw, Poland
\item \Idef{org140}Wayne State University, Detroit, Michigan, United States
\item \Idef{org141}Westf\"{a}lische Wilhelms-Universit\"{a}t M\"{u}nster, Institut f\"{u}r Kernphysik, M\"{u}nster, Germany
\item \Idef{org142}Wigner Research Centre for Physics, Hungarian Academy of Sciences, Budapest, Hungary
\item \Idef{org143}Yale University, New Haven, Connecticut, United States
\item \Idef{org144}Yonsei University, Seoul, Republic of Korea
\end{Authlist}
\endgroup
\end{document}